%% file: main.tex
\documentclass[acmsmall]{acmart}

\acmJournal{TOSEM}

\usepackage[utf8]{inputenc}
\usepackage[T1]{fontenc}

\usepackage{newtxmath} 
\usepackage{array}
\usepackage{graphicx}
\usepackage{subfig}
\usepackage[ruled,vlined,lined,commentsnumbered]{algorithm2e}
\usepackage{balance}
\usepackage{algorithmic}
\usepackage{esvect}
\usepackage{booktabs}
\usepackage{tabularx}
\usepackage{anyfontsize}
\usepackage[all]{nowidow}
\usepackage[normalem]{ulem}
\usepackage{xcolor}
\usepackage{colortbl}
\usepackage{enumitem}
\usepackage{makecell}
\usepackage[switch]{lineno}

\newcolumntype{Y}{>{\centering\arraybackslash}X}

\definecolor{commentGray}{RGB}{120,120,120}
\renewcommand{\algorithmiccomment}[1]{\bgroup\color{commentGray}{//#1}\egroup}

\newcommand\algorithmicinput{\textbf{Input:~}}
\newcommand\INPUT{\item[\algorithmicinput]}
\newcommand\algorithmicoutput{\textbf{Output:~}}
\newcommand\OUTPUT{\item[\algorithmicoutput]}

\newcommand\algorithmichinput{\quad\quad\quad}
\newcommand\HINPUT{\item[\algorithmichinput]}
\newcommand\algorithmichoutput{\quad\quad\quad}
\newcommand\HOUTPUT{\item[\algorithmichoutput]\hspace{0.75em}}

\newcommand{\algorithmicbreak}{\textbf{break}}
\newcommand{\BREAK}{\STATE \algorithmicbreak}

\usepackage{framed}
\usepackage{tikz}
\definecolor{light-gray}{gray}{0.9}
\usepackage[framemethod=TikZ]{mdframed}
\mdfsetup{skipabove=5pt,skipbelow=3pt}
\usepackage{lipsum}
\mdfdefinestyle{RQFrame}{%
	linecolor=black,
	outerlinewidth=0.15pt,
	roundcorner=3pt,
	innertopmargin=2pt,
	innerbottommargin=2pt,
	innerrightmargin=4pt,
	innerleftmargin=4pt,
	backgroundcolor=light-gray}

\begin{document}
\title{Estimating Probabilistic Safe WCET Ranges of Real-Time Systems at Design Stages}
\input{authors}
\input{tex/abstract}

\setcounter{page}{1}

\maketitle

\input{tex/intro}
\input{tex/motivation}
\input{tex/probdesc}
\input{tex/approach}
\input{tex/casestudy}

\input{tex/relatedwork}
\input{tex/conclusion}
\input{tex/acknowledgments}

\bibliographystyle{ACM-Reference-Format}
\balance
\bibliography{bib/ref}


\end{document}

%% file: authors.tex
\author{Jaekwon Lee}
\email{jaekwon.lee@uni.lu}
\affiliation{%
  \institution{University of Luxembourg}
  \streetaddress{29 Avenue John F. Kennedy}
  \city{Luxembourg}
  \postcode{1859}
  \country{Luxembourg}
}
\affiliation{%
  \institution{University of Ottawa}
  \streetaddress{800 King Edward Avenue}
  \city{Ottawa}
  \postcode{ON K1N 6N5}  
  \country{Canada}
}

\author{Seung Yeob Shin}
\email{seungyeob.shin@uni.lu}
\affiliation{%
  \institution{University of Luxembourg}
  \streetaddress{29 Avenue John F. Kennedy}
  \city{Luxembourg}
  \postcode{1859}
  \country{Luxembourg}
}

\author{Shiva Nejati}
\email{snejati@uottawa.ca}
\affiliation{%
  \institution{University of Ottawa}
  \streetaddress{800 King Edward Avenue}
  \city{Ottawa}
  \postcode{ON K1N 6N5}  
  \country{Canada}
}
\affiliation{%
  \institution{University of Luxembourg}
  \streetaddress{29 Avenue John F. Kennedy}
  \city{Luxembourg}
  \postcode{1859}  
  \country{Luxembourg}
}

\author{Lionel C. Briand}
\email{lionel.briand@uni.lu}
\affiliation{%
  \institution{University of Luxembourg}
  \streetaddress{29 Avenue John F. Kennedy}
  \city{Luxembourg}
  \postcode{1859}  
  \country{Luxembourg}
}
\affiliation{%
  \institution{University of Ottawa}
  \streetaddress{800 King Edward Avenue}
  \city{Ottawa}
  \postcode{ON K1N 6N5}  
  \country{Canada}
}

\author{Yago Isasi Parache}
\email{isasi@luxspace.lu}
\affiliation{%
  \institution{LuxSpace}
  \streetaddress{9 Rue Pierre Werner}
  \city{Betzdorf}
  \postcode{6832}  
  \country{Luxembourg}
}

%% file: tex/abstract.tex
\begin{abstract}
Estimating worst-case execution times (WCET) is an important activity at early design stages of real-time systems. Based on WCET estimates, engineers make design and implementation decisions to ensure that task executions always complete before their specified deadlines. However, in practice, engineers often cannot provide precise point WCET estimates and prefer to provide plausible WCET ranges. Given a set of real-time tasks with such ranges, we provide an automated technique to determine for what WCET values the system is likely to meet its deadlines, and hence operate safely with a probabilistic guarantee. Our approach combines a search algorithm for generating worst-case scheduling scenarios with polynomial logistic regression for inferring probabilistic safe WCET ranges. We evaluated our approach by applying it to three industrial systems from different domains and several synthetic systems. Our approach efficiently and accurately estimates probabilistic safe WCET ranges within which deadlines are likely to be satisfied with a high degree of confidence.

\end{abstract}

\ccsdesc[500]{Software and its engineering~Search-based software engineering}
\ccsdesc[500]{Computer systems organization~Real-time systems}

\ccsdesc[500]{Software and its engineering~Search-based software engineering}

\keywords{Schedulability Analysis, Worst-Case Execution Time, Meta-Heuristic Search, Machine Learning, Search-Based Software Engineering}

%% file: tex/intro.tex
\section{Introduction}
\label{sec:intro}

Safety-critical systems, e.g., those used in the aerospace, automotive and healthcare domains, require that their executions always complete before their specified deadlines in all execution scenarios, including the worst cases. The systems that must perform their operations in such a timely manner are known as real-time systems (RTS)~\cite{Burns:09}. To ensure that a real-time system meets its deadlines, we need an accurate estimation of the worst-case execution times (WCET) of software tasks that concurrently run in the system. For instance, the Anti-lock Braking System (ABS) of a vehicle has to activate within milliseconds after the driver brakes. However, an ABS taking more time for activation than the estimated WCET may result in a vehicle skid due to the wheels locking up.

Accurately estimating WCET values of a real-time system is particularly important at early design stages when real-time tasks are not yet fully implemented. Accurate WCET estimates greatly support engineers during development as they provide targets driving design and implementation choices. Based on the WCET estimates, engineers can make design and implementation decisions, e.g., using either a relational database or an in-memory data storage. In addition, when engineers develop software and hardware components in parallel, which is common in the aerospace, automotive, and healthcare domains, accurately estimated WCET values at early design stages help engineers find optimal configurations of hardware devices, e.g., CPUs, sensors, and actuators, by accounting for time constraints and performance requirements of the system.

Real-time tasks have various parameters such as task priorities, deadlines, inter-arrival times, and WCET values~\cite{Stankovic:98,Cottet:02}. Among the task parameters, WCET values are typically difficult to accurately estimate at early design stages. The other parameters, however, can be specified or estimated with a high degree of precision even at early stages. For example, task priorities are typically determined by the selected scheduling policy, e.g., rate monotonic~\cite{Liu:73}, or based on the task criticality levels (i.e., more critical tasks are prioritised over the less critical ones). Task deadlines are typically decided by system requirements. Task inter-arrival times, i.e., the time interval between consecutive task executions, usually depend on system environmental events triggering task executions. In contrast, the WCET values of some tasks may depend on various factors such as implementation decisions, task durations, real-time operating systems, and hardware components. However, these factors may not be fully known at early stages of development, making it difficult to precisely estimate WCET values for real-time tasks~\cite{Gustafsson:09,Altenbernd:16,Bonenfant:17}. As a result, engineers tend to provide ranges for WCET values instead of point estimates.

The problem of estimating WCET values is, in general, a hard problem. WCET values of real-time tasks impact every possible task schedules. WCET values depend on the content and implementation of tasks and not the schedule. But they impact how the tasks are scheduled. The space of all possible task schedules is very large. In our context, the problem becomes computationally more expensive when WCET values are uncertain and are specified as value ranges instead of single values. Specifically, provided with WCET value ranges, engineers need to have ways to determine for what WCET values, within the given ranges, the system is likely to miss or satisfy its deadline constraints. If engineers know that deadlines are likely met for all or most of the expected WCET ranges, they can consider a wider choice of design and implementation options. Otherwise, in situations where only tight WCET sub-ranges seem acceptable, developers may have to consider more expensive hardware, decreased functionality or performance, or more restricted design and implementation choices.

The WCET estimation problem of real-time systems has been widely studied in the past~\cite{Bernat:02,Gustafsson:09,Hansen:09,Altenbernd:16,Hardy:16,Bonenfant:17}. To our knowledge, however, most of the existing WCET estimation methods often fail to provide WCET estimates at early design stages because they require inputs that can be defined only at a later point in time such as task implementations and hardware devices. Some model-based approaches~\cite{Cimatti:08,Sun:13,Andre:19} try to solve the WCET estimation problem exhaustively by applying a model checker to a real-time model, e.g., parametric timed automata, of the system under analysis. These exhaustive methods are applicable once real-time system models are available. However, such approaches tend to suffer from the state-space explosion problem~\cite{Clarke:12} as the number of software tasks and their different states increase. More recently, stress testing and simulation-based approaches~\cite{Briand:05,Alesio:13} have been proposed to stress RTS and generate test scenarios where their deadline constraints are violated. Such approaches cast the schedulability test problem as an optimisation problem to find worst-case task execution scenarios exhibiting deadline misses. However, none of the existing simulation-based approaches account for uncertainties in WCET values and therefore do not handle WCET value ranges. Our work complements the simulation-based stress testing approach and extends it to account for uncertainties in WCET values.


\textbf{Contributions.} In this article, we propose a \underline{S}afe WCET \underline{A}nalysis method \underline{F}or real-time task sch\underline{E}dulability (SAFE) to estimate WCET ranges under which tasks are likely to be schedulable with a probabilistic guarantee. Our approach is based on a stress testing approach~\cite{Briand:05} using meta-heuristic search~\cite{Luke:13} in combination with polynomial logistic regression models. Specifically, we use a genetic algorithm~\cite{Luke:13} to search for sequences of task arrivals which likely lead to deadline misses. Then, logistic regression~\cite{David:86}, a statistical classification technique, is applied to infer a \emph{safe WCET border} in the multidimensional WCET space with a probabilistic guarantee. This border aims to partition the given WCET ranges into \emph{safe} and \emph{unsafe} sub-ranges for a selected deadline miss probability $p$, and thus enables engineers to investigate trade-offs among different tasks' WCET values. WCET ranges are deemed to be probabilistically safe if tasks, within such ranges, have a high probability to complete their executions before their specified deadlines. In this article, for the sake of simplicity, we refer to probabilistically safe WCET ranges as safe WCET ranges. We evaluated our approach by applying it to a complex, industrial satellite system developed by our industry partner, LuxSpace, as well as two industrial systems from different domains and several synthetic systems. Results show that our approach can efficiently and accurately compute safe WCET ranges. SAFE scales to complex industrial systems as an offline analysis method. Execution times of SAFE on our industrial systems are practically acceptable, i.e., at most 27h.
To our knowledge, SAFE is the first attempt to estimate safe WCET ranges within which real-time tasks are likely to meet their deadlines for a given level of confidence, while enabling engineers to explore trade-offs among tasks' WCET values. Our full evaluation package is available online~\cite{Artifacts}.

\textbf{Organization.} The remainder of this article is structured as follows: Section~\ref{sec:motivation} motivates our work. Section~\ref{sec:problem} defines our specific schedulability analysis problem in practical terms. Section~\ref{sec:approach} describes SAFE. Section~\ref{sec:eval} evaluates SAFE. Sections~\ref{sec:relatedwork} compares SAFE with related work. Section~\ref{sec:conclusion} concludes this article.

%% file: tex/motivation.tex
\section{Motivating case study}
\label{sec:motivation}

We motivate our work with a mission-critical real-time satellite system, named Attitude Determination and Control System (ADCS), which LuxSpace, a leading system integrator for microsatellites and aerospace systems, has been developing over the years. ADCS determines the satellite's attitude and controls its movements~\cite{Eickhoff:11}. ADCS controls a satellite in either autonomous or passive mode. In the autonomous mode, ADCS must orient a satellite in proper position on time to ensure that the satellite provides normal service correctly. In the passive mode, operators are able to not only control satellite positions but also maintain the satellite, e.g., upgrading software. Such a maintenance operation does not necessarily need to be completed within a fixed hard deadline; instead, it should be completed within a reasonable amount of time, i.e., soft deadlines. Hence, ADCS is composed of a set of tasks having real-time constraints with hard and soft deadlines.


Engineers at LuxSpace conduct real-time schedulability analysis across different development stages. At an early design stage, when task implementations and system hardware are not available, the engineers use a theoretical schedulability analysis technique~\cite{Liu:73} which determines that a set of tasks is schedulable if CPU utilisation of the task set is less than a threshold, e.g., 69\%. As mentioned earlier, at an early design stage, engineers estimate task WCETs as ranges and often assign large values to the upper bounds of such ranges. To be on the safe side, engineers tend indeed to be conservative in their analysis.

Engineers, however, are still faced with the following issues: (1)~An analytical schedulability analysis technique, e.g., utilisation-based schedulability analysis~\cite{Liu:73}, typically indicates whether or not tasks are schedulable. However, engineers need additional information to understand how tasks miss their deadlines. For instance, a set of tasks may not be schedulable for a few specific sequences of task arrivals. (2)~Engineers estimate WCETs without any systematic support; instead, they often rely on their experience of developing tasks providing similar functions-to-develop. This practice typically results in imprecise estimates of WCET ranges, which may cause serious problems, e.g., significantly changing tasks at later development stages. To this end, LuxSpace is interested in SAFE as a way to address these issues in analysing schedulability.

%% file: tex/probdesc.tex
\section{Problem description}
\label{sec:problem}

This section first formalises task, task relationship, scheduler, and schedulability concepts. We then describe the problem of identifying safe WCET ranges under which tasks likely meet their deadline constraints at a certain level of confidence; i.e., tasks are schedulable with a certain probability.

\textbf{Task.} A real-time system is composed of a set of $n$ tasks that should complete their executions within specified deadlines after they are activated (or arrived). We denote by $\tau_i$ a real-time task indexed by $i$ in the range from $1$ to $n$. Every real-time task $\tau_i$ has the following properties: priority denoted by $P_i$, deadline denoted by $D_i$, and worst-case execution time (WCET) denoted by $C_i$. Task priority $P_i$ determines if an execution of a task is preempted by another task. Typically, a task $\tau_i$ preempts the execution of a task $\tau_j$ if the priority of $\tau_i$ is higher than the priority of $\tau_j$, i.e., $P_i > P_j$.

The deadline of a task $\tau_i$ relative to its arrival time is denoted by $D_i$. A task deadline can be either \emph{hard} or \emph{soft}. A hard deadline of a task $\tau_i$ specifies that $\tau_i$ \emph{must} complete its execution within a deadline $D_i$ after $\tau_i$ is activated. While violations of hard deadlines are not acceptable, depending on the operating context of a system, violating soft deadlines may be tolerated to some extent. Note that, for notational simplicity, we do not introduce new notations to distinguish between hard and soft deadlines. In this article, we refer to a hard deadline as a deadline. Section~\ref{sec:approach} further discusses how our approach manages hard and soft deadlines.

We denote by $C^{min}_i$ and $C^{max}_i$, respectively, the minimum and the maximum WCET values of a task $\tau_i$. As discussed in the introduction, at an early development stage, it is difficult to provide exact WCET values of real-time tasks. Hence, we assume that engineers specify WCETs using a range of values, instead of single values, by indicating estimated minimum and maximum values that they think each task's WCET can realistically take.

In this article, real-time tasks are either \emph{periodic} or \emph{aperiodic}. Periodic tasks, which are typically triggered by timed events, are invoked at regular intervals specified by their \emph{period}. We denote by $T_i$ the period of a periodic task $\tau_i$, i.e., a fixed time interval between subsequent activations (or arrivals) of $\tau_i$. Aperiodic tasks have irregular arrival times and are activated by external stimuli which occur irregularly, and hence, in general, there is no limit on the arrival times of an aperiodic task. However, in real-time analysis, we typically specify a minimum inter-arrival time denoted by $T^{min}_i$ and a maximum inter-arrival time denoted by $T^{max}_i$ indicating the minimum and maximum time intervals between two consecutive arrivals of an aperiodic task $\tau_i$. In real-time analysis, sporadic tasks are often separately defined as having irregular arrival intervals and hard deadlines~\cite{Liu:00}. In our conceptual definitions, however, we do not introduce new notations for sporadic tasks because the deadline and period concepts defined above are sufficient to characterise sporadic tasks. Note that for a periodic task $\tau_i$, we have $T^{min}_i = T^{max}_i = T_i$. Otherwise, for an aperiodic task $\tau_j$, we have $T^{max}_j > T^{min}_j$.

\textbf{Task relationship.} The execution of a task $\tau_i$ depends not only on its own parameters described above, e.g., priority $P_i$ and period $T_i$, but also on its relationships with other tasks. Relationships between tasks are typically determined by task interactions related to accessing shared resources~\cite{Alesio:12}, such as memory, file, and IO devices. Specifically, if two tasks $\tau_i$ and $\tau_j$ access a shared resource in a mutually exclusive way, $\tau_i$ may be blocked from executing for the period during which $\tau_j$ accesses the resource. We denote by $\mathit{dp}(\tau_i,\tau_j)$ the resource-dependency relation between tasks $\tau_i$ and $\tau_j$ that holds if $\tau_i$ and $\tau_j$ have mutually exclusive access to a shared resource such that they cannot be executed in parallel or preempt each other, but one can execute only after the other has completed its access to the resource. We note that  resource-dependency relations are defined at the level of tasks, following prior works~\cite{Locke:90,Anssi:11,Alesio:15} describing the industrial case study systems used in our experiments (see Section~\ref{subsec:casestudy}). The $\mathit{dp}(\tau_i,\tau_j)$ relation is symmetric, i.e., $\mathit{dp}(\tau_i,\tau_j)$ = $\mathit{dp}(\tau_j,\tau_i)$.

\textbf{Scheduler.} Let $\Gamma$ be a set of tasks to be scheduled by a real-time scheduler. A scheduler then dynamically schedules executions of tasks in $\Gamma$ according to the tasks' arrivals and the scheduler's scheduling policy over the scheduling period $\mathbb{T} = [0,\mathbf{t}]$. We denote by $a_{i,k}$ the $k$th arrival time of a task $\tau_i \in \Gamma$. The first arrival of a periodic task $\tau_i$ does not always occur immediately at the system start time $0$. Such offset time from the system start time $0$ to the first arrival time $a_{i,1}$ of $\tau_i$ is denoted by $O_i$. For a periodic task $\tau_i$, the $k$th arrival of $\tau_i$ within $\mathbb{T}$ is $a_{i,k} \leq \mathbf{t}$ and is computed by $a_{i,k} = O_i + (k-1) \cdot T_i$. For an aperiodic task $\tau_j$, $a_{j,k}$ is determined based on the $k{-}1$th arrival time of $\tau_j$ and its minimum and maximum arrival times. Specifically, for $k > 1$, $a_{j,k} \in [a_{j,{k-1}}+T^{min}_{j}, a_{j,{k-1}}+T^{max}_{j}]$ and, for $k = 1$, $a_{j,1} \in [T^{min}_{j}, T^{max}_{j}]$ where $a_{j,k} \le \mathbf{t}$.

A scheduler reacts to a task arrival at $a_{i,k}$ to schedule the execution of $\tau_i$. Depending on a scheduling policy (e.g., rate monotonic scheduling policy~\cite{Liu:73} and single-queue multi-core scheduling policy~\cite{Arpaci:18}), an arrived task $\tau_i$ may not start its execution at the same time as it arrives when a higher priority task is executing. Also, task executions may be interrupted due to preemption. We denote by $e_{i,k}$ the end execution time for the $k$th arrival of a task $\tau_i$. Depending on actual worst-case execution time of a task $\tau_i$, denoted by $C_i$, within its WCET range $[C^{min}_i, C^{max}_i]$, the $e_{i,k}$ end execution time of $\tau_i$ satisfies the following: $e_{i,k} \ge a_{i,k} + C_i$.

During the system operation, a scheduler generates a \emph{schedule scenario} which describes a sequence of task arrivals and their end time values. We define a schedule scenario as a set $S$ of tuples $(\tau_i, a_{i,k}, e_{i,k})$ indicating that a task $\tau_i$ has arrived at $a_{i,k}$ and completed its execution at $e_{i,k}$. Due to the randomness of task execution times and aperiodic task arrivals, a scheduler may generate a different schedule scenario in different runs of a system.

\begin{figure}[t]
\begin{center}
\subfloat[Deadline miss]
{\parbox{1\columnwidth}{\centering
\includegraphics[width=0.9\columnwidth]{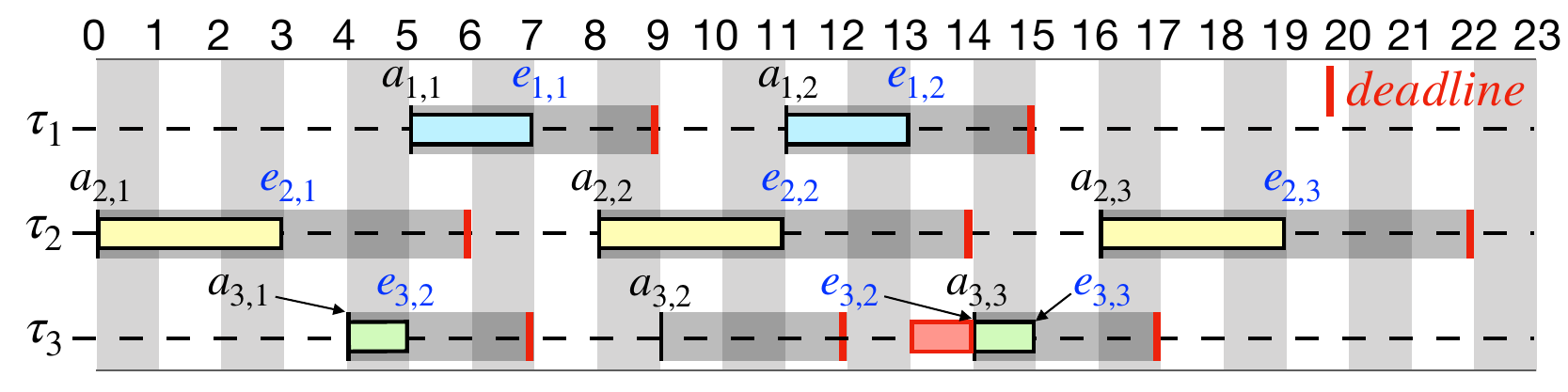}\label{fig:dm}
}}%

\subfloat[No deadline miss]
{\parbox{1\columnwidth}{\centering
\includegraphics[width=0.9\columnwidth]{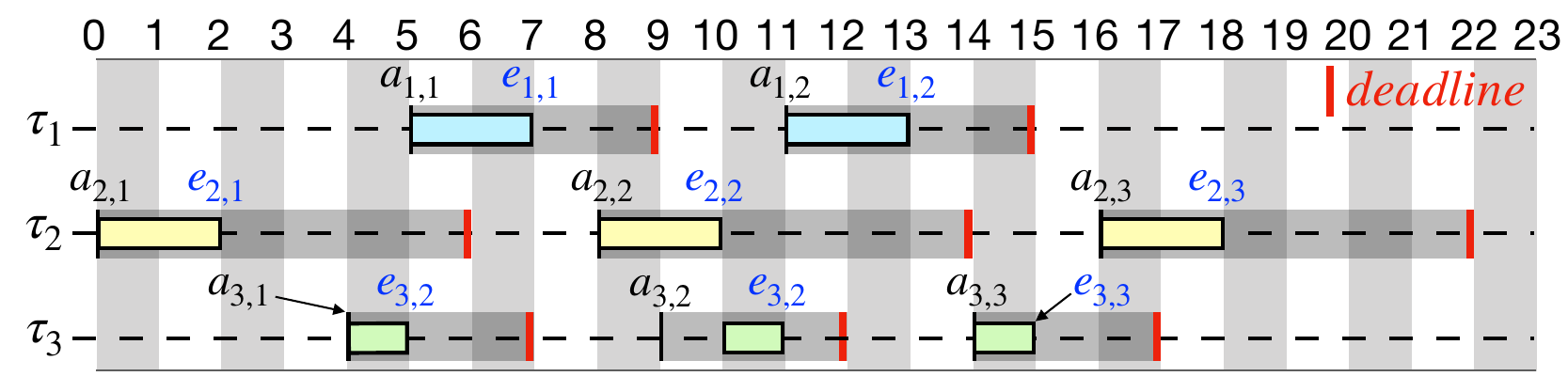}\label{fig:nodm}
}}%
\caption{Example schedule scenarios of three tasks, $\tau_1$, $\tau_2$, and $\tau_3$, running on a single core system. (a)~$\tau_3$ is not schedulable, i.e., $e_{3,2} > a_{3,2} + D_3$. (b)~All the three tasks are schedulable. When $\tau_2$ executes over 3 (WCET) time units, it causes a deadline miss of $\tau_3$. When the WCET is reduced to 2, the three tasks are schedulable even for the same sequence of task arrivals.}
\label{fig:scheduling}
\end{center}
\end{figure}

Figure~\ref{fig:scheduling} shows two schedule scenarios produced by a scheduler over the $[0,23]$ time period of a system run. Both Figure~\ref{fig:dm} and Figure~\ref{fig:nodm} describe executions of three tasks, $\tau_1$, $\tau_2$, and $\tau_3$ arrived at the same time stamps (see $a_{i,k}$ in the figures). In both scenarios, the aperiodic task $\tau_1$ is characterised by: $T^{min}_{1} = 5$, $T^{max}_{1} = 10$, $D_1 = 4$, and $C^{min}_1 = C^{max}_1 = 2$. The periodic task $\tau_2$ is characterised by: $T_2 = 8$ and $D_2 = 6$. The aperiodic task $\tau_3$ is characterised by: $T^{min}_3 = 3$, $T^{max}_3 = 20$, $D_3 = 3$, and $C^{min}_3 = C^{max}_3 = 1$. The priorities of the three tasks satisfy the following: $P_1 > P_2 > P_3$. In both scenarios, task executions can be preempted depending on their priorities. We note that a WCET range of the $\tau_2$ task is set to $C^{min}_2 = 1$ and $C^{max}_2 = 3$ in Figure~\ref{fig:dm}, and $C^{min}_2 = 1$ and $C^{max}_2 = 2$ in Figure~\ref{fig:nodm}. Then, Figure~\ref{fig:dm} can be described by the $S_a = \{(\tau_1, 5, 7)$, $\ldots$, $(\tau_2, 0, 3)$, $\ldots$, $(\tau_3, 9, 14)$, $(\tau_3, 14, 15)\}$ schedule scenario; and Figure~\ref{fig:nodm} by $S_b = \{(\tau_1, 5, 7)$, $\ldots$, $(\tau_2, 0, 2)$, $\ldots$, $(\tau_3, 9, 11)$, $(\tau_3, 14, 15)\}$.

\textbf{Schedulability.} Given a schedule scenario $S$, a task $\tau_i$ is \emph{schedulable} if $\tau_i$ completes its execution before its deadline, i.e., for all $e_{i,k}$ observed in $S$, $e_{i,k} \le a_{i,k} + D_i$. Let $\Gamma$ be a set of tasks to be scheduled by a scheduler. A set $\Gamma$ of tasks is then schedulable if for every schedule $S$ of $\Gamma$, we have no task $\tau_i \in \Gamma$ that misses its deadline. 

As shown in Figure~\ref{fig:dm}, a deadline miss occurs after the second arrival of $\tau_3$, i.e., $e_{3,2} > a_{3,2} + D_3$. During $[a_{3,2}, a_{3,2} + D_3]$ period, the $\tau_3$ task cannot execute because the other tasks $\tau_1$ and $\tau_2$ with higher priorities are executing. Thus, $\tau_3$ is not schedulable in the schedule scenario of Figure~\ref{fig:dm}. This scheduling problem can be solved by restricting tasks' WCET ranges  as discussed below.

\textbf{Problem.} Uncertainty in task WCET values at an early development stage is a critical issue preventing the effective design and assessment of mission-critical real-time systems. Upper bounds of WCETs correspond to worst-case WCET values and have a direct impact on deadline misses as larger WCET values increase their probability. Lower bounds of WCETs are estimates of tasks' best-case WCET values, below which task implementations are likely not feasible. Our approach aims to determine the maximum upper bounds for WCET under which tasks are likely to be schedulable, at a given level of risk, and thus provides an objective to engineers implementing the tasks. Specifically, for every task $\tau_i \in \Gamma$ to be analysed, our approach computes a new upper bound value for the WCET range of $\tau_i$ (denoted by $C^\mathit{max*}_i$) such that  $C^\mathit{max*}_i \leq C^{max}_i$ and by restricting the WCET range of $\tau_i$ to $C^\mathit{max*}_i$ we should, at a certain level of confidence, no longer have deadline misses. That is, tasks $\Gamma$ become schedulable, with a certain probability, after restricting the maximum WCET value of $\tau_i$ to $C^\mathit{max*}_i$. For instance, as shown in Figure~\ref{fig:nodm}, restricting the maximum WCET of $\tau_2$ from $C^{max}_2 = 3$ to $C^{max*}_2 = 2$ enables all the three tasks to be schedulable.

We note that, in our context, both arrival time ranges for aperiodic tasks and WCET ranges for all tasks are represented as continuous intervals. Since our approach works based on sampling values from these continuous ranges, our approach cannot be exhaustive and cannot provide a guarantee that the tasks can always be schedulable after restricting their WCET ranges. Our approach instead relies on sampling values within the WCET and arrival time ranges, simulating the scheduler behaviour using the sampled values and observing whether, or not, a deadline miss occurs. In lieu of exhaustiveness, we rely on statistical and machine learning techniques to provide probabilistic estimates indicating how confident we are that a given set of tasks are schedulable. 



%% file: tex/approach.tex
\section{Approach}\label{sec:approach}

\begin{figure}[t]
\begin{center}
\includegraphics[width=0.8\columnwidth]{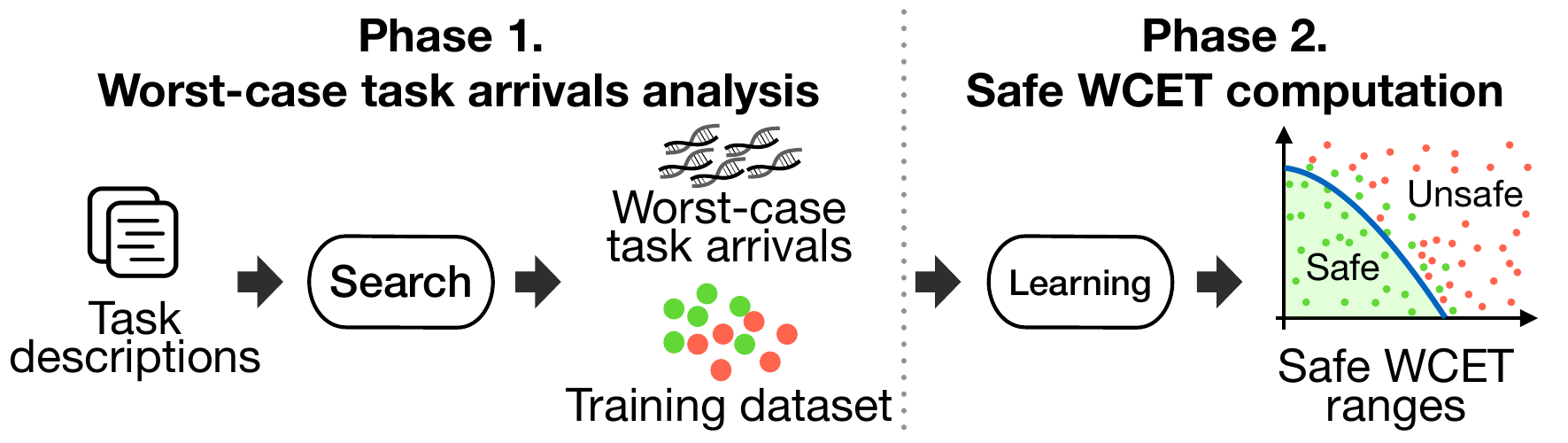}
\caption{An overview of our \underline{S}afe WCET \underline{A}nalysis method \underline{F}or real-time task sch\underline{E}dulability (SAFE).}
\label{fig:over}
\end{center}
\end{figure}

Figure~\ref{fig:over} shows an overview of our \underline{S}afe WCET \underline{A}nalysis method \underline{F}or real-time task sch\underline{E}dulability (SAFE). Phase 1 of SAFE aims at searching worst-case task-arrival sequences. A task-arrival sequence is worst-case if deadline misses are maximised or, when this is not possible, tasks complete their executions as close to their deadlines as possible. Building on existing work, we identify worst-case task-arrival sequences using a search-based approach relying on genetic algorithms. Phase 2 of SAFE, which is the main contribution of this article, aims at computing safe WCET ranges under which tasks are likely to be schedulable. To do so, relying on logistic regression and an effective sampling strategy, we augment the worst-case task-arrival sequences generated in Phase 1  to compute safe WCET ranges with a certain deadline miss probability, indicating a degree of risk. We describe in detail \hbox{these two phases next.} 

\subsection{Phase 1: Worst-case task arrivals}
\label{subsec:phase1}

The first phase of SAFE finds worst-case sequences in the space of possible sequences of task arrivals, defined by their inter-arrival time characteristics. As SAFE aims to provide conservative, safe WCET ranges, we optimise task arrivals to maximise task completion times and deadline misses, and indirectly minimise safe WCET ranges (see the safe area visually presented in Figure~\ref{fig:over}). 
We address this optimisation problem using a single-objective search algorithm. Following standard practice~\cite{Ferrucci:13}, we describe our search-based approach for identifying worst-case task arrivals by defining the solution representation, the scheduler, the fitness function, and the computational search algorithm. We then describe the dataset of sequences generated by search and then used for training our logistic regression model to compute safe WCET ranges in the second phase of SAFE.

Our approach in Phase 1 is based on past work~\cite{Briand:05}, where a specific genetic algorithm configuration was proposed to find worst-case task arrival sequences.  One important modification though is that we account for uncertainty in WCET values through simulations for evaluating the magnitude of deadline misses.

\textbf{Representation.} Given a set $\Gamma$ of tasks to be scheduled, a feasible solution is a set $A$ of tuples $(\tau_i, a_{i,k})$ where $\tau_i \in \Gamma$ and $a_{i,k}$ is the $k$th arrival time of a task $\tau_i$. Thus, a solution $A$ represents a valid sequence of task arrivals of $\Gamma$ (see valid $a_{i,k}$ computation in Section~\ref{sec:problem}). Let $\mathbb{T} = [0, \mathbf{t}]$ be the time period during which a scheduler receives task arrivals. The size of $A$ is equal to the number of task arrivals over the $\mathbb{T}$ time period. Due to the varying inter-arrival times of aperiodic tasks (Section~\ref{sec:problem}), the size of $A$ will vary across different solutions.


\textbf{Scheduler.} SAFE uses a simulation technique for analysing the schedulability of tasks to account for the uncertainty in WCET values and  scalability issues. For instance, an inter-arrival time of a software update task in a satellite system is approximately at most three months. In such cases, conducting an analysis based on an actual scheduler is prohibitively expensive. Instead, SAFE uses a real-time task scheduling simulator, named SafeScheduler, which samples WCET values from their ranges for simulating task executions and applies a scheduling policy, i.e., single-queue multi-core scheduling policy~\cite{Arpaci:18}, based on discrete simulation time events. Note that we chose the single-queue multi-core scheduling policy for SafeScheduler since our case study systems (described in Section~\ref{subsec:casestudy}) rely on this policy.

SafeScheduler takes a feasible solution $A$ for scheduling a set $\Gamma$ of tasks as an input. It then outputs a schedule scenario as a set $S$ of tuples $(\tau_i,a_{i,k},e_{i,k})$ where $a_{i,k}$ and $e_{i,k}$ are the $k$th arrival and end time values of a task $\tau_i$, respectively. Recall from Section~\ref{sec:problem} that SafeScheduler computes task arrivals based on periodic tasks' offsets and periods and aperiodic tasks' inter-arrival times. For each task $\tau_i$, SafeScheduler computes $e_{i,k}$ based on its scheduling policy and a selected WCET value for $\tau_i$ within the WCET range $[C^{min}_i, C^{max}_i ]$, while accounting for resource-dependency relationships (see Section~\ref{sec:problem}). Hence, each run of SafeScheduler for the same input solution $A$ will likely produce a different schedule scenario.

SafeScheduler implements a single-queue multi-core scheduling policy~\citep{Arpaci:18}, which schedules a task $\tau_i$ with explicit priority $P_i$ and deadline $D_i$. When tasks arrive, SafeScheduler puts them into a single queue that contains tasks to be scheduled. At any simulation time, if there are tasks in the queue and multiple cores are available to execute tasks, SafeScheduler first fetches a task $\tau_i$ from the queue in which $\tau_i$ has the highest priority $P_i$. SafeScheduler then allocates task $\tau_i$ to any available core. Note that if task $\tau_i$ shares a resource with a running task $\tau_j$ in another core, i.e., the $\mathit{dp}(\tau_i,\tau_j)$ resource-dependency relationship holds, SafeScheduler follows standard task-blocking rules~\cite{Liu:00}, i.e., $\tau_i$ will be blocked until $\tau_j$ releases the shared resource.

SafeScheduler works under the assumption that context switching time is free, which is also a working assumption in many scheduling analysis methods~\citep{Liu:73,Audsley:01,Alesio:15}. Note that the assumptions are practically valid and useful at an early development step in the context of real-time analysis. For instance, our collaborating partner accounts for the waiting time of tasks due to context switching between tasks through adding some extra time to WCET ranges at the task design stage. Note that SAFE can be applied with any scheduling policy, including those that account for context switching time and multiple queues.

\textbf{Fitness.} Given a feasible solution $A$ for a set $\Gamma$ of tasks, we formulate a fitness function, $f(A,\Gamma^\delta,ns)$, to quantify the degree of deadline misses regarding a set $\Gamma^\delta \subseteq \Gamma$ of target tasks, where $ns$ is a number of SafeScheduler runs to account for the uncertainty in WCET. SAFE provides the capability of selecting target tasks $\Gamma^\delta$ as practitioners often need to focus on the most critical tasks. We denote by $\mathit{dist}(\tau_i,k)$ the distance between the end time and the deadline of the $k$th arrival of task $\tau_i$  and define $\mathit{dist}(\tau_i,k) = e_{i,k} - a_{i,k} + D_i$ (see Section~\ref{sec:problem} for the notation end time $e_{i,k}$, arrival time $a_{i,k}$, and deadline $D_i$).

To compute the  $f(A,\Gamma^\delta,ns)$ fitness value, SAFE runs SafeScheduler $ns$ times for $A$ and obtains ${ns}$ schedule scenarios $S_1, S_2, \ldots, S_{ns}$. For each schedule scenario $S_h$, we denote by $\mathit{dist}_h({\tau_i, k})$ the distance between the end and deadline time values corresponding to the $k$th arrival of the task $\tau_i$ observed in $S_h$. We denote by $\mathit{lk}(\tau_i)$ the last arrival index of a task $\tau_i$ in $A$. SAFE aims to maximise the $f(A,\Gamma^\delta,ns)$ fitness function defined as follows:
\begin{equation*}
f(A,\Gamma^\delta,ns) = \sum_{h =1}^{ns}\max_{\tau_i \in \Gamma^\delta,~ k \in [1,\mathit{lk}(\tau_i)]}\mathit{dist}_h(\tau_i, k) / {ns}
\end{equation*}

We note that soft deadline tasks also require to execute within reasonable execution time ranges. Hence, engineers also estimate safe WCET ranges for soft deadline tasks. As the above fitness function returns a quantified degree of deadline misses, SAFE uses such function for both soft and hard deadline tasks.

\textbf{Computational search.} SAFE employs a steady-state genetic algorithm~\cite{Luke:13}. The algorithm breeds a new population for the next generation after computing the fitness of a population. The breeding for generating the next population is done by using the following genetic operators: (1)~\emph{Selection.} SAFE selects candidate solutions using a tournament selection technique, with the tournament size equal to two which is the most common setting~\cite{Gendreau:10}. (2)~\emph{Crossover.} Selected candidate solutions serve as parents to create offspring using a crossover operation. (3)~\emph{Mutation.} The offspring are then mutated. Below, we describe our crossover and mutation operators.


\emph{Crossover.} A crossover operator is used to produce offspring by mixing traits of parent solutions. SAFE modifies the standard one-point crossover operator~\cite{Luke:13} as two parent solutions $A_p$ and $A_q$ may have different sizes, i.e., $|A_p| \neq |A_q|$. Let $\Gamma = \{\tau_1, \tau_2, \ldots, \tau_n\}$ be a set of tasks to be scheduled. Our crossover operator, named SafeCrossover, first randomly selects an aperiodic task $\tau_i \in \Gamma$. 
For all $j \in [1,i]$ and $\tau_j \in \Gamma$, SafeCrossover then swaps all $\tau_j$ arrivals between two solutions $A_p$ and $A_q$. As the size of $\Gamma$ is fixed for all solutions, SafeCrossover can cross over two solutions that may have different sizes.

\begin{table}[t]
\caption{An example operation of SafeCrossover. It swaps all task arrivals of task $\tau_1$ and $\tau_2$ between two parent solutions $A_p$ and $A_q$ to produce offspring $A^\prime_p$ and $A^\prime_q$.}
\label{tbl:crossover example}
\centering
\begin{tabularx}{\columnwidth}{XXXX}
\toprule
& Task $\tau_1$ & Task $\tau_2$ & Task $\tau_3$ \\ 
\midrule
Parent $A_p$ & \cellcolor{light-gray}{$(\tau_1,5)$, $(\tau_1,11)$} & \cellcolor{light-gray}{$(\tau_2,8)$, $(\tau_2,16)$} & $(\tau_3,4)$, $(\tau_3,10)$\\
Parent $A_q$ & $(\tau_1,3)$, $(\tau_1,7)$, $(\tau_1,14)$ & $(\tau_2,6)$, $(\tau_2,13)$ & \cellcolor{light-gray}{$(\tau_3,5)$, $(\tau_3,8)$, $(\tau_3,13)$}\\
Child $A^\prime_p$ & $(\tau_1,3)$, $(\tau_1,7)$, $(\tau_1,14)$ & $(\tau_2,6)$, $(\tau_2,13)$ & $(\tau_3,4)$, $(\tau_3,10)$ \\
Child $A^\prime_q$ & \cellcolor{light-gray}{$(\tau_1,5)$, $(\tau_1,11)$} & \cellcolor{light-gray}{$(\tau_2,8)$, $(\tau_2,16)$} & \cellcolor{light-gray}{$(\tau_3,5)$, $(\tau_3,8)$, $(\tau_3,13)$} \\
\bottomrule
\end{tabularx}
\end{table}

Table~\ref{tbl:crossover example} shows an example operation of SafeCrossover using a system with three aperiodic tasks, $\tau_1$, $\tau_2$, and $\tau_3$. Let two parent solutions $A_p$ and $A_q$ be as follows: $A_p$ $=$ $\{(\tau_1,5)$, $\ldots$, $(\tau_2,8)$, $\ldots$, $(\tau_3,10)\}$ and $A_q$ $=$ $\{(\tau_1,3)$, $\ldots$, $(\tau_2,6)$, $\ldots$, $(\tau_3,13)\}$, where $(\tau_i,t)$ denotes task $\tau_i$ arrives at time $t$. Given the two parents $A_p$ and $A_q$, SafeScheduler randomly selects a task, i.e., $\tau_2$ in this example, then it swaps all arrivals of $\tau_1$ and $\tau_2$ between $A_p$ and $A_q$. As shown in Table~\ref{tbl:crossover example}, SafeCrossover then generates the offspring $A^\prime_p$ and $A^\prime_q$ as follows: $A^\prime_p$ $=$ $\{(\tau_1,3)$, $\ldots$, $(\tau_2,6)$, $\ldots$, $(\tau_3,10)\}$ and $A^\prime_q$ $=$ $\{(\tau_1,5)$, $\ldots$, $(\tau_2,8)$, $\ldots$, $(\tau_3,13)\}$. The shaded (resp. unshaded) cells in Table~\ref{tbl:crossover example} indicate which task arrivals in child $A^\prime_q$ (resp. $A^\prime_p$) come from which parent.

\emph{Mutation operator} SAFE uses a heuristic mutation algorithm, named SafeMutation. For a solution $A$, SafeMutation mutates the $k$th task arrival time $a_{i,k}$ of an aperiodic task $\tau_i$ with a mutation probability.
SafeMutation chooses a new arrival time value of $a_{i,k}$ based on the $[T^{min}_i, T^{max}_i]$ inter-arrival time range of $\tau_i$. If such a mutation of the $k$th arrival time of $\tau_i$ does not affect the validity of the $k{+}1$th arrival time of $\tau_i$, the mutation operation ends. Specifically, let $a^*_{i,k}$ be a mutated value of $a_{i,k}$. In case $a_{i,k+1} \in [a^*_{i,k} + T^{min}_i, a^*_{i,k} + T^{max}_i]$, SafeMutation returns the mutated $A$ solution.

After mutating the $k$th arrival time $a_{i,k}$ of a task $\tau_i$ in a solution $A$, if the $k{+}1$th arrival becomes invalid, SafeMutation corrects the remaining arrivals of $\tau_i$. We denote by $a^*_{i,k}$ the mutated $k$th arrival time of $\tau_i$. For all the arrivals of $\tau_i$ after $a^*_{i,k}$, SafeMutation first updates their original arrival time values by adding the difference $a^*_{i,k}-a_{i,k}$. Let $\mathbb{T} = [0,\mathbf{t}]$ be the scheduling period. SafeMutation then removes some arrivals of $\tau_i$ if they are mutated to arrive after $\mathbf{t}$ or adds new arrivals of $\tau_i$ while ensuring that all tasks arrive within $\mathbb{T}$.

Given the offspring presented in Table~\ref{tbl:crossover example}, SafeMutation, for example, mutates a child solution $A^\prime_p$ $=$ $\{(\tau_1,3)$, $(\tau_1,7)$, $(\tau_1,14)$, $\ldots$, $(\tau_3,10)\}$. Let $[T^{min}_1, T^{max}_1]$ $=$ $[2,8]$ be the inter-arrival time range of task $\tau_1$, $\mathbb{T} = [0,18]$ be the time period during which SafeScheduler receives task arrivals, and SafeMutation selects the second arrival of task $\tau_1$, i.e., $(\tau_1,7)$ in Table~\ref{tbl:crossover example}, to mutate. Based on the inter-arrival time range of $\tau_1$, SafeMutation randomly chooses a new arrival time, e.g., $5$, for the second arrival of $\tau_1$. The third arrival $(\tau_1,14)$ of $\tau_1$ then became invalid due to the mutated second arrival $(\tau_1,5)$; i.e., $\tau_1$ cannot arrive at time $14$ because $14$ $\notin$ $[5 + 2, 5 + 8]$, where $[T^{min}_1, T^{max}_1]$ $=$ $[2,8]$. According to the correction procedure described above, the third arrival of $\tau_1$ is modified to $(\tau_1,12)$ as $12$ $=$ $14 + (5-7)$, where $14$, $5$, and $7$ are, respectively, the original thrid arrival time of $\tau_1$, the original second arrival time of $\tau_1$, and the mutated second arrival time of $\tau_1$. As SafeScheduler can receive new arrivals of $\tau_1$ after time $12$, SafeMutation may add new arrivals of $\tau_1$ based on the inter-arrival time range of $\tau_1$.

We note that when a system is only composed of periodic tasks, SAFE will skip searching for worst-case arrival sequences as arrivals of periodic tasks are deterministic (see Section~\ref{sec:problem}), but will nevertheless generate the labelled dataset described below. When needed, SAFE can be easily extended to manipulate varying offset (and period) values for periodic tasks, in a way identical to how we currently handle inter-arrival times.


\textbf{Labelled dataset.} SAFE infers safe WCET ranges using a supervised learning technique~\cite{Russell:10} which requires a labelled dataset, namely logistic regression.
In our context, a supervised learning technique creates a model that correlates tasks' WCET values with schedulability results indicating whether these tasks meet their deadlines or not. Supervised learning is conducted based on pairs of tasks' WCET values and a schedulability result, i.e., a labelled dataset. Specifically, SAFE uses logistic regression because it allows engineers to have probabilistic interpretation of safe WCET ranges and to investigate trade-off relationships among different tasks' WCETs. Section~\ref{subsec:phase2} describes this learning process in detail.

Recall from the fitness computation described above, SAFE runs SafeScheduler $ns$ times to obtain schedule scenarios $S {=} {\{S_1, S_2,\ldots,S_{ns}\}}$, and then computes a fitness value of a solution $A$ based on $S$. We denote by $W_h$ a set of tuples $(\tau_i,C_i)$ representing that a task $\tau_i$ has the $C_i$ WCET value in the $S_h$ schedule scenario. Let $\vv{L}$ be a labelled dataset to be created by the first phase of SAFE. We denote by $\ell_h$ a label indicating whether or not a schedule scenario $S_h$ has any deadline miss for any of the target tasks in $\Gamma^\delta$, i.e., $\ell_h$ is either $\mathit{safe}$ or $\mathit{unsafe}$ which denotes, respectively, no deadline miss or deadline miss. For each fitness computation, SAFE adds $ns$ number of tuples $(W_h,\ell_h)$ to $\vv{L}$. Specifically, for a schedule scenario $S_h$, SAFE adds $(W_h,\mathit{unsafe})$ to $\vv{L}$ if there are $\tau_i {\in} \Gamma^\delta$ and $k {\in} [1, \mathit{lk}(i)]$ such that $\mathit{dist}_h(\tau_i,k) {>} 0$; otherwise SAFE \hbox{adds $(W_h, \mathit{safe})$ to $\vv{L}$.}

\subsection{Phase 2: Safe ranges of WCET}
\label{subsec:phase2}

\begingroup
\begin{algorithm}[t]
\parbox{0.95\columnwidth}{\caption{SafeRefinement. An algorithm for computing safe WCET ranges under which target tasks are schedulable. The algorithm consists of three steps as follows: ``reduce complexity'', ``handle imbalanced dataset'', and ``refine model'' steps.}\label{alg:safewcet}}
\SetKwFunction{ReduceDimension}{ReduceDimension}
\SetKwFunction{LearnRegressionModel}{LearnRegressionModel}
\SetKwFunction{StepwiseRegression}{StepwiseRegression}
\SetKwFunction{Regression}{Regression}
\SetKwFunction{Probability}{Probability}
\SetKwFunction{HandleImbalance}{HandleImbalance}
\SetKwFunction{RunSafeScheduler}{RunSafeScheduler}
\SetKwFunction{Add}{Add}
\SetKwFunction{Sample}{SampleWCET}
\SetKwFunction{Precision}{PrecisionByCrossValidation}

\begin{algorithmic}[1]
\INPUT- $\vv{L}$: Labelled dataset obtained from the SAFE search
\HINPUT- $G$: Worst solutions obtained from the SAFE search
\HINPUT- $\mathit{ns}$: Number of WCET samples per solution
\HINPUT- $\mathit{nl}$: Number of logistic regression models
\HINPUT- $\mathit{pt}$: Precision threshold
\OUTPUT- $m$: Safe WCET model
\HOUTPUT- $p$: Probability of deadline misses
\BlankLine
\STATE \COMMENT {step 1. reduce complexity}
\STATE $\vv{L}^r$ $\leftarrow$ \ReduceDimension {$\vv{L}$} \COMMENT {feature reduction}
\STATE $m$ $\leftarrow$ \StepwiseRegression {$\vv{L}^r$} \COMMENT {term selection}
\STATE $p$ $\leftarrow$ \Probability {$m$, $\vv{L}^r$}
\STATE \COMMENT {step 2. handle imbalanced dataset}
\STATE $\vv{L}^b$ $\leftarrow$ \HandleImbalance {$\vv{L}^r$, $m$}
\STATE \COMMENT {step 3. refine model}
\FOR {$\mathit{nl}$ times}
  \STATE \COMMENT {step 3.1. add new data instances}
  \FOR {each solution $A \in G$}
    \STATE $\{S_1, S_2, \ldots, S_{\mathit{ns}}\}$ $\leftarrow$ \RunSafeScheduler {$A$, $m$, $p$, $\mathit{ns}$}
    \FOR {each scenario $S_h \in \{S_1, S_2, \ldots, S_{\mathit{ns}}\}$}
		  \IF {$S_h$ has any deadline miss}
		    \STATE $\vv{L}^b$ $\leftarrow$ \Add {$\vv{L}^b, (\mathit{WCET}(S_h),\mathit{unsafe})$}
		  \ELSE
		    \STATE $\vv{L}^b$ $\leftarrow$ \Add {$\vv{L}^b, (\mathit{WCET}(S_h),\mathit{safe})$}
		  \ENDIF
    \ENDFOR
  \ENDFOR
  \STATE \COMMENT {step 3.2. learn regression model}
  \STATE $m$ $\leftarrow$ \Regression {$m$, $\vv{L}^b$}
  \STATE $p$ $\leftarrow$ \Probability {$m$, $\vv{L}^b$}
 	\IF {\Precision {$m$,$\vv{L}^b$} $> \mathit{pt}$}
 		\BREAK
  \ENDIF
\ENDFOR
\RETURN $m$, $p$
\end{algorithmic}
\end{algorithm}
\endgroup

In Phase 2, SAFE computes safe ranges of WCET values under which target tasks are likely to be  schedulable. To do so, SAFE applies a supervised machine learning technique to the labelled dataset generated by Phase 1 (Section~\ref{subsec:phase1}). Specifically, Phase 2 executes SafeRefinement (Algorithm~\ref{alg:safewcet}) which has following steps: complexity reduction, imbalance handling and model refinement.

\textbf{Complexity reduction.} The ``reduce complexity'' step in Algorithm~\ref{alg:safewcet} reduces the dimensionality of a labelled dataset $\vv{L}$ obtained from the first phase of SAFE (line 2). It predicts initial safe WCET ranges based on the WCET variables for the tasks in $\Gamma$ (line 3) that have the most significant effect on deadline misses for target tasks. A labelled dataset $\vv{L}$ obtained from the first phase of SAFE contains tuples $(W,\ell)$ where $W$ is a set of WCET values for tasks in $\Gamma$ and $\ell$ is a label of $W$ indicating either no deadline miss (safe) or deadline miss (unsafe) (Section~\ref{subsec:phase1}). Note that some WCET values in $W$ may not be relevant to determine $\ell$. Hence, $\vv{L}$ may contain irrelevant variables to predict $\ell$. To decrease computational complexity for the remaining steps, SafeRefinement creates a reduced dataset $\vv{L}^r$ which contains the same number of data instances (tuples) as $\vv{L}$ while including only WCET values with a significant effect on $\ell$. To that end, SafeRefinement employs a standard feature reduction technique: random forest feature reduction~\cite{Breiman:01} which has been successfully applied to high-dimensional data~\cite{Nguyen:15,Hideko:12}. Given the labelled dataset $\vv{L}$, random forest creates a set of decision trees based on the parameter values such as the number of trees and tree depth. Decision trees obtained by random forest allow us to rank features, i.e., task WCETs, based on their importance as measured by Gini impurity~\cite{Breiman:01}. Hence, by setting a particular threshold for importance, we can select a subset of the features. Note that Section~\ref{subsec:impl} describes the parameter values for the feature reduction step in detail.

\begin{figure}[t]
\begin{center}
\includegraphics[width=0.7\linewidth]{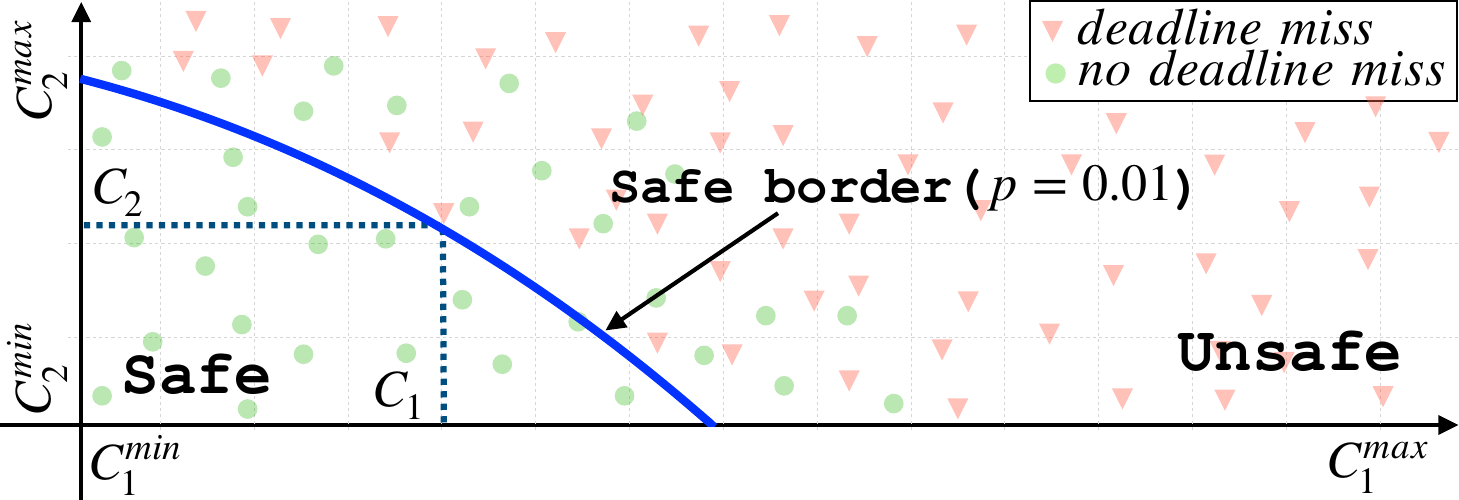}
\caption{A safe border line of WCET values for the $\tau_1$ and $\tau_2$ tasks. The safe border is determined by a deadline miss probability of 0.01. $C_1$ and $C_2$ determine safe WCET ranges of $\tau_1$ and $\tau_2$ under which they likely satisfy their deadlines.}
\label{fig:wspace}
\end{center}
\end{figure}

After reducing the dimensionality of the input dataset $\vv{L}$ in Algorithm~\ref{alg:safewcet}, resulting in the reduced dataset $\vv{L}^r$, SafeRefinement learns an initial model to predict safe WCET ranges. SafeRefinement uses logistic regression~\cite{Hosmer:13} because it enables a probabilistic interpretation of safe WCET ranges and the investigation of relationships among different tasks' WCETs. For example, Figure~\ref{fig:wspace} shows a \emph{safe border} determined by an inferred logistic regression model $m$ with a probability $p$ of deadline misses. Note that a safe range, e.g., $[C^{min}_1, C_1]$ of task $\tau_1$ in Figure~\ref{fig:wspace}, is determined by a point on the safe border in a multidimensional WCET space.
A safe border distinguishes safe and unsafe areas in the WCET space. After inferring a logistic regression model $m$ from the input dataset, SafeRefinement selects a probability $p$ maximising the safe area under the safe border determined by $m$ and $p$ while ensuring that all the data instances, i.e., sets of WCET values, classified as safe using the safe border are actually observed to be safe in the input dataset, i.e., no false positives (lines 3--4).
We note that engineers can also select an adequate probability, which may yield false positives or not maximise the area under the safe border, depending on their needs.

SafeRefinement uses a second-order polynomial response surface model (RSM)~\cite{Munda:08} to build a logistic regression model. RSM is known to be useful when the relationship between several explanatory variables (e.g., WCET variables) and one or more response variables (e.g., safe or unsafe label) needs to be investigated~\cite{Munda:08,Muhuri:18}. RSM contains linear terms, quadratic terms, and 2-way interactions between linear terms. 
Let $V$ be a set of WCET variables $v_x$ in $\vv{L}^r$. Then, the logistic regression model of SafeRefinement is defined as follows:
\begin{equation*}
\log \frac{p}{1-p} = c_0 + \sum_{x=1}^{|V|}{c_xv_x} + \sum_{x=1}^{|V|}{c_{xx}v_x^2} + \sum_{y>x}{c_{xy}v_xv_y}
\end{equation*}
As shown in the above equation, an RSM equation, i.e., the right-hand side, built on the reduced dataset $\vv{L}^r$ has a higher number of dimensions, i.e., the number of coefficients to be inferred, than $|V|$ as RSM additionally accounts for quadratic terms ($v_x^2$) and 2-way interactions ($v_xv_y$) between linear terms. Hence, SafeRefinement employs a stepwise regression technique (line 3), e.g., stepwise AIC (Akaike Information Criterion)~\cite{Yamashita:07}, in order to select significant explanatory terms from the RSM equation. This allows the remaining ``refine model'' step of SafeRefinement to execute efficiently as it requires to run SafeScheduler and logistic regression multiple times within a time budget (line 8), both operations being computationally expensive.

\begin{figure}[t]
\begin{center}
\includegraphics[width=0.7\linewidth]{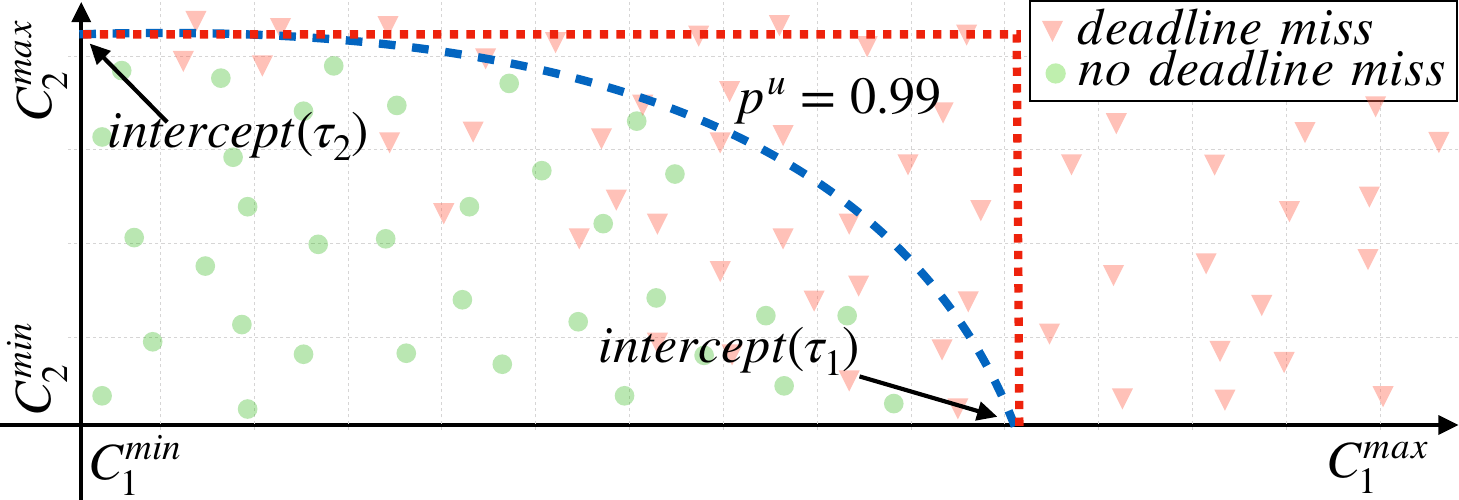}
\caption{Handling imbalanced dataset by excluding unsafe WCET values based on logistic regression intercepts.}
\label{fig:imbalance}
\end{center}
\end{figure}

\textbf{Imbalance handling.} Recall from Section~\ref{subsec:phase1} that SAFE searches for worst-case sequences of task arrivals and is guided by maximising the magnitude of deadline misses, when they are possible. Therefore, the major portion of $\vv{L}$, the dataset produced by the first phase of SAFE, is a set of task arrival sequences leading to deadline misses. Supervised machine learning techniques (including logistic regression) typically produce unsatisfactory results when faced with highly imbalanced datasets~\cite{Batista:04}. SafeRefinement addresses this problem with the ``handle imbalanced dataset'' step in Algorithm~\ref{alg:safewcet} (lines 5--6) before refining safe WCET ranges. SafeRefinement aims to identify WCET ranges under which tasks are likely to be schedulable. This entails that WCET ranges under which tasks are highly unlikely to be schedulable can be safely excluded from the remaining analysis.  Specifically, SafeRefinement prunes out WCET ranges with a high probability of deadline misses above a high threshold $p^u$ and thus creates a more balanced dataset $\vv{L}^b$ compared to the original imbalanced dataset $\vv{L}^r$ (line 6). SafeRefinement automatically finds a minimum probability $p^u$ which leads to a safe border classifying no false unsafe (negative) instances in $\vv{L}^r$. SafeRefinement then updates the maximum WCET $C^{max}_i$ of a task $\tau_i$ based on the intercept of the logistic regression model $m$ (with a probability of $p^u$) on the WCET axis for $\tau_i$. Figure~\ref{fig:imbalance} shows an example dataset $\vv{L}^r$ with a safe border characterised by a high deadline miss probability, i.e., $p^u = 0.99$, to create a more balanced dataset $\vv{L}^b$ within the restricted ranges $[C^{min}_1, \mathit{intercept}(\tau_1)]$ and $[C^{min}_2, \mathit{intercept}(\tau_2)]$.

\textbf{Model refinement.} The ``refine model'' step in Algorithm~\ref{alg:safewcet} refines an inferred logistic regression model by sampling additional schedule scenarios selected according to a strategy that is expected to improve the model. As described in Section~\ref{subsec:phase1}, the SAFE search produces a set $G$ (population) of worst-case arrival sequences of tasks $\Gamma$ which likely violate deadline constraints of target tasks $\Gamma^\delta \subseteq \Gamma$. For each arrival sequence $A$ in $G$, SafeRefinement executes SafeScheduler $\mathit{ns}$ times to add $\mathit{ns}$ new data instances to the dataset $\vv{L}^b$ based on the generated schedule scenarios and their schedulability results (lines 9--19). After adding $\mathit{ns} \cdot |G|$ new data instances to $\vv{L}^b$, SafeRefinement runs logistic regression again to infer a refined logistic regression model $m$ and computes a probability $p$ that ensures no false safe instances (positives) in $\vv{L}^b$ and maximises the safe area under the safe border defined by $m$ and $p$ (lines 20--25).

In the second phase of SAFE, SafeScheduler selects WCET values for tasks in $\Gamma$ to compute a schedule scenario based on a distance-based random number generator, which extends the standard uniform random number generator. The distance-based WCET value sampling aims at minimising the Euclidean distance between the sampled WCET points and the safe border defined by the inferred model $m$ and the selected probability $p$. SafeScheduler iteratively computes new WCET values using the following distance-based sampling procedure: (1)~generating $nr$ random samples in the WCET space, (2)~computing their distance values from the safe border, and (3)~selecting the closest point to the safe border.

SafeRefinement stops model refinements either by reaching an allotted analysis budget (line 8 of Algorithm~\ref{alg:safewcet}) or when a precision reaches an acceptable level $\mathit{pt}$, e.g., 0.99 (lines 23--25). SafeRefinement uses the standard precision metric~\cite{Witten:11} as described in Section~\ref{subsec:metrics}. In our context, practitioners need to identify safe WCET ranges at a high level of precision to ensure that identified safe WCET ranges can be trusted. To compute a precision value, SafeRefinement uses a standard k-fold cross-validation~\cite{Witten:11}. In k-fold cross-validation, $\vv{L}^b$ is partitioned into k equal-size splits. One split is retained as a test dataset, and the remaining k-1 splits are used as a training dataset. The cross-validation process is then repeated k times to compute a precision of inferred safe borders which are determined by a logistic regression model $m$ and a probability $p$ (lines 21 and 22)


\textbf{Selecting WCET ranges.} A safe border defined by an inferred logistic regression model and a deadline miss probability of $p$ represents a (possibly infinite) set of points, corresponding to safe WCET ranges of tasks, e.g., $[C^{min}_1, C_1]$ and $[C^{min}_2, C_2]$ in Figure~\ref{fig:wspace}. In practice, however, engineers need to choose a specific WCET range for each task to conduct further analysis and development. How to choose optimal WCET ranges depends on the system context. At  early  stages, however, such contextual information may not be available. Hence, SAFE proposes a \emph{best-size point}, i.e., WCET ranges, on a safe border which maximises the volume of the hyperbox the point defines. In general, the larger hyperbox, the greater flexibility the engineers have in selecting appropriate WCET values. Choosing the point with the largest volume is helpful when no domain-specific information is available to define other selection criteria. In general the inferred safe border enables engineers to investigate trade-off  among different tasks' WCET values.

%% file: tex/casestudy.tex
\section{Evaluation}
\label{sec:eval}

We evaluate SAFE using an industrial case study from the satellite domain. Our full evaluation package is available online~\cite{Artifacts}.

\subsection{Research Questions}
\label{subsec:rq}


\noindent\textbf{RQ1 (baseline comparison):} \textit{How does SAFE perform compared with a baseline approach?} With RQ1, we investigate whether SAFE can outperform WCET estimation based on random search. Note that such RQ is an important \emph{sanity check} for search-based solutions in general~\cite{Harman:12,Arcuri:14}. Our conjecture is that SAFE, although computationally expensive, will significantly outperform a random search solution with respect to estimating safe WCET ranges with a higher degree of confidence.

\noindent\textbf{RQ2 (effectiveness of distance-based sampling):} \textit{How does SAFE, based on distance-based sampling, perform compared with random sampling?} We compare our distance-based sampling procedure described in Section~\ref{subsec:phase2} and used in the second phase of SAFE with a naive random sampling. Our conjecture is that distance-based sampling, although expensive, is needed to improve the quality of the training data used for logistic regression. RQ2 assesses this conjecture by comparing distance-based and random sampling. 

\noindent\textbf{RQ3 (usefulness):} \textit{Can SAFE identify WCET ranges within which tasks are highly likely to satisfy their deadline constraints?} In RQ3, we investigate whether SAFE identifies acceptably safe WCET ranges in practical time. We further discuss our insights regarding the usefulness of SAFE from the feedback obtained from engineers in LuxSpace.

\noindent\textbf{RQ4 (scalability):} \textit{Can SAFE find safe WCET ranges for large-scale systems with a practical time budget?} In this RQ, we study the relationship between the execution time of SAFE and the parameters of study subjects. We use several synthetic subjects to be able to freely control key real-time systems' parameters.

\subsection{Industrial Study Subjects}
\label{subsec:casestudy}

\begin{table*}[t]
\caption{Description of the three industrial subject systems: number of periodic and aperiodic tasks, resource dependencies, and platform cores. The full task descriptions are available online~\cite{Artifacts}.}
\label{tbl:subjects}
\begin{center}
\begin{tabularx}{\columnwidth}{XXXXX}
\toprule
System & Periodic tasks & Aperiodic tasks & Dependencies & Cores \\
\midrule
ADCS & 15 & 19 & 0 & 1 \\
ICS  &  3 &  3 & 3 & 3 \\
UAV  & 12 &  4 & 4 & 3 \\
\bottomrule
\end{tabularx}
\end{center}
\end{table*}

We evaluated SAFE by applying it to our motivating case study subject, i.e., the satellite attitude determination and control system (ADCS) described in Section~\ref{sec:motivation}, as well as two industrial study subjects from the literature~\cite{Frati:08,Traore:06}. Table~\ref{tbl:subjects} summarizes the relevant attributes of these subjects, presenting the number of periodic and aperiodic tasks, resource dependencies, and processing cores. The subjects are characterized by real-time parameters, e.g., priorities, WCETs, periods and deadlines, described in Section~\ref{sec:problem}. The full task descriptions of the subjects are available online~\cite{Artifacts}. The main missions of the three subjects are describe as follows:
\begin{itemize}[leftmargin=1em]
\item ADCS is a satellite system that aims at orienting a satellite in a proper position on time to ensure that the satellite provides normal service correctly (see Section~\ref{sec:motivation}). LuxSpace, our industry partner, developed ADCS for an ESA project.
\item ICS is an ignition control system that checks the status of an automotive engine and corrects any errors of the engine~\cite{Frati:08}. The system was developed by Bosch GmbH\footnote{Bosch {GmbH}: https://www.bosch.com/}.
\item UAV is a mini unmanned air vehicle that follows dynamically defined way-points and communicates with a ground station to receive instructions~\cite{Traore:06}. The system was developed in a collaboration with the University of Poitiers France and ENSMA\footnote{ENSMA: https://www.ensma.fr/}.
\end{itemize}

LuxSpace is a leading system integrator of micro satellites and aerospace systems. ADCS includes a set of 15 periodic and 19 aperiodic tasks. Eight tasks out of the 19 aperiodic tasks are constrained by hard deadlines, i.e., sporadic tasks. Out of the 34 tasks, engineers provided single WCET values for eight tasks. For the remaining 26 tasks, engineers estimated WCET ranges due to uncertain decisions, e.g., implementation choices and hardware specifications, made at later development stages (see Section~\ref{sec:motivation}). The differences between the estimated WCET maximum and minimum values across the 26 tasks varies from 0.1ms to 20000ms. Our collaboration with LuxSpace enabled us to discuss SAFE results with engineers to draw important qualitative conclusions and to assess the benefits of SAFE (see Section~\ref{subsec:res}).

For the experiments with ICS and UAV, we used the task descriptions reported in a previous study~\cite{Alesio:15} and modified their tasks' WCETs from point values to ranges. Though the problem of schedulability analysis of real-time tasks has been widely studied~\cite{Alesio:15,Altenbernd:16,Hardy:16,Bonenfant:17,Bruggen:18}, none of the prior work addresses the same problem (see Section~\ref{sec:problem}) as that addressed by SAFE. Hence, the public study subjects in the literature do not fit our study's requirements. In particular, none of the public real-time system case studies~\cite{Alesio:15} contains estimated WCET ranges in their task descriptions. These ranges, however, are necessary to apply SAFE and to evaluate its effectiveness. In order to evaluate SAFE in various and realistic system contexts, we chose to apply SAFE to existing industrial subjects, i.e., ICS and UAV, described in prior work~\cite{Alesio:15} and made necessary changes only to task WCETs of the subjects as described below. Compared to ADCS, ICS and UAV have different task characteristics, such as resource dependencies and number of processing cores.

We note that estimating (practically valid) WCET ranges requires significant domain expertise. For public domain case study systems such as ICS and UAV, however, we do not have any access to the engineers who have developed those subjects. Hence, we chose to apply a simple and straightforward method to convert a point WCET value to a WCET range as follows: (Step 1) We first check whether the system under analysis is schedulable or not. For a task $\tau_i$ in the system, we denote by $C_i$ an original point WCET value of $\tau_i$. (Step 2) If the system is evaluated to be schedulable, it indicates that the system's tasks may be able to handle higher execution times than their estimated WCETs.  Hence, we simply define the WCET range of $\tau_i$ by $[C_i,~r{\cdot}C_i]$, where $r>1$, as input WCET ranges for SAFE. This modification enables SAFE to find more relaxed safe WCET ranges. (Step 2$^\prime$) Otherwise, if the system is evaluated to be unschedulable, we define the WCET range of $\tau_i$ by $[r^\prime{\cdot}C_i,~C_i]$, where $r^\prime<1$, as input for SAFE. This modification allows SAFE to find appropriate WCET estimates, ensuring the system is likely to be schedulable under the WCET ranges found by SAFE. As ICS and UAV are likely to be schedulable~\cite{Alesio:15}, for all task $\tau_i$ in ICS and task $\tau_j$ in UAV, we created the modified task descriptions of ICS and UAV based on Step 2. We conducted experiments using simulations to set the $r$ values for ICS and UAV by configuring $r$ to 1.1, 1.2, $\ldots$, 1.5 incrementally until we could find deadline misses in each system, i.e., unsafe WCET values. Recall from Section~\ref{subsec:phase2} that SAFE relies on logistic regression to partition the given WCET ranges into safe and unsafe sub-ranges for a selected deadline miss probability. Hence, we modified the estimated WCET ranges of ICS and UAV to include both safe and unsafe WCET ranges.
Given the experiment results, we set the WCET ranges of ICS and UAV to $[C_i,1.2{\cdot}C_i]$ and $[C_j,1.5{\cdot}C_j]$, respectively.
The full original and modified task descriptions of ICS and UAV are available online~\cite{Artifacts}.

\subsection{Synthetic study subjects}
\label{subsec:synthetic}

\begingroup
\begin{algorithm}[t]
\parbox{0.95\columnwidth}{
    \caption{An algorithm for creating a synthetic subject while accounting for the task characteristics described in Section~\ref{sec:problem}.}
    \label{alg:synthetic}
}
\SetKwFunction{genTaskset}{generate\_task\_set}
\SetKwFunction{UUniFast}{UUniFast\_discard}
\SetKwFunction{genOffset}{generate\_offsets}
\SetKwFunction{genTasks}{generate\_task\_periods}
\SetKwFunction{convAperiodics}{convert\_to\_aperiodic\_tasks}
\SetKwFunction{convRangedWCET}{convert\_to\_WCET\_ranges}
\SetKwFunction{checkValidity}{is\_valid}
   
\begin{algorithmic}[1]
\INPUT- ${n}$: number of tasks
\HINPUT- $u^t$: target utilisation
\HINPUT- $T^\mathit{min}$: minimum task period
\HINPUT- $T^\mathit{max}$: maximum task period
\HINPUT- $g$: granularity of task periods
\HINPUT- $\theta$: maximum offset value
\HINPUT- $\gamma$: ratio of aperiodic tasks
\HINPUT- $\mu$: range factor to determine inter-arrival times
\HINPUT- $\omega$: number of WCET ranges
\HINPUT- $\lambda$: range factor to determine WCET ranges
\OUTPUT- $\Gamma$: set of tasks
\BlankLine

\STATE $\Gamma \leftarrow$ $\{\}$, $\mathbf{C} \leftarrow$ $\{\}$
\STATE \COMMENT {synthesise a set of periodic tasks}
\STATE $\mathbf{U} \leftarrow$ \UUniFast{$n, u^t$} 
\STATE $\mathbf{T} \leftarrow$ \genTasks{$n, T^\mathit{min}, T^\mathit{max}, g$} \COMMENT {task periods}

\FOR {each $i \in [1,n]$}
    \STATE $\mathbf{C}$ $\leftarrow$ $\mathbf{C} \cup \{U_i {\cdot} T_i\}$, where $U_i \in \mathbf{U}$ and $T_i \in \mathbf{T}$ \COMMENT {WCETs}
\ENDFOR
\STATE $ \Gamma \leftarrow$ \genTaskset{$\mathbf{T}, \mathbf{C}$, $\theta$, $g$} 

\STATE \COMMENT {convert some periodic tasks to aperiodic tasks}
\STATE $ \Gamma \leftarrow$ \convAperiodics{$\Gamma, \gamma, \mu$}
\STATE \COMMENT {convert some WCET point values to WCET ranges}
\STATE $ \Gamma \leftarrow$ \convRangedWCET{$\Gamma, \omega,
\lambda$}
\RETURN $\Gamma$
\end{algorithmic}
\end{algorithm}
\endgroup

We evaluated the scalability of SAFE using synthetic systems, following the common scalability analysis practice applied in many real-time system studies~\cite{Davis:08,Zhang:09,Davis:11a,Grass:18,Vonder:18,Durr:19}. As shown in Algorithm~\ref{alg:synthetic}, we synthesise a set of real-time tasks by varying key task parameters as described below. The algorithm first synthesises a set of periodic tasks (lines 2-8) and then converts some of these tasks to aperiodic tasks (lines 9-10). Last, the algorithm configures some tasks with WCET ranges (lines 11-12).

As shown on line 3 of Algorithm~\ref{alg:synthetic}, the algorithm first creates a set $\mathbf{U}$ of task utilisation values by using the UUniFast-Discard algorithm~\cite{Davis:11a} that is devised to give an unbiased distribution of task utilisation values. The UUniFast-Discard algorithm takes as input the number of tasks to be synthesised, $n$, and a target utilisation value, $u^t$. It then outputs $n$ utilization values, $U_1$ $\ldots$ $U_n$, where $0 < U_i < 1$ for all $U_i$ and $\sum_{i=1}^{n}{U_i} = u^t$.

As for line 4 in Algorithm~\ref{alg:synthetic}, the algorithm generates $n$ task periods, $T_1$ $\ldots$ $T_n$ according to a log-uniform distribution within a range [$T^{min}, T^{max}$], i.e., given a task period (random variable) $T_i$, $\mathit{log}~T_i$ follows a uniform distribution. For example, when a period range [$T^{min}, T^{max}$] is [10ms, 1000ms], the algorithm generates approximately an equal number of tasks in period ranges [10ms, 100ms] and [100ms, 1000ms]. The parameter $g$ is used to determine the granularity of period values as multiples of $g$. Lines 5-7 of Algorithm~\ref{alg:synthetic} describe how the algorithm synthesises tasks' WCET values. Specifically, for each task $\tau_i$, the algorithm computes the WCET value $C_i$ of $\tau_i$ as $C_i = U_i \cdot T_i$.

Given the task periods $\mathbf{T}$ and the WCET values $\mathbf{C}$, line 8 of Algorithm~\ref{alg:synthetic} synthesizes a set $\Gamma$ of periodic tasks accounting for offsets, priorities, and deadlines. A periodic task $\tau_i$ is characterised by a period $T_i$, a WCET $C_i$, an offset $O_i$, a priority $P_i$, and a deadline $D_i$ (see Section~\ref{sec:problem}). A task offset $O_i$ is randomly selected from an input range $[0, \theta]$ of offset values. The algorithm relies on the rate-monotonic scheduling policy~\cite{Liu:73} to decide task priorities and deadlines. Specifically, tasks with shorter periods are given higher priorities and tasks' deadlines are equal to their periods.

Line 10 of Algorithm~\ref{alg:synthetic} synthesises aperiodic tasks. The algorithm converts some periodic tasks into aperiodic tasks according to a ratio $\gamma$ of aperiodic tasks among all tasks. The algorithm then uses a range factor $\mu$ to determine minimum and maximum inter-arrival times of aperiodic tasks. Specifically, for a task $\tau_i$ to be converted, the algorithm computes a range  $[T^{min}_i, T^{max}_i]$ of inter-arrival times as  $[T^{min}_i, T^{max}_i]$ $=$ $[T_i \times (1-\mu), T_i \times (1+\mu)]$, where $\mu \in (0,1)$. For example, if $\mu=0.45$ and $T_i=50$ for a task $\tau_i$ to be converted, $[T^{min}_i, T^{max}_i]$ $=$ $[27.5, 72.5]$.

To synthesise tasks' WCET ranges, line 12 of Algorithm~\ref{alg:synthetic} randomly selects $\omega$ tasks in $\Gamma$ to convert their WCET point values to WCET ranges. For a selected task $\tau_i$, the algorithm computes a WCET range $[C^{min}_i, C^{max}_i]$ as $[C^{min}_i, C^{max}_i]$ $=$ $[C_i \times (1-\lambda), C_i \times (1+\lambda)]$, where $\lambda$ is a range factor to determine the WCET ranges and $\lambda \in (0,1)$. For example, if $\lambda=0.25$ and $C_i=10$ for a task $\tau_i$, $[C^{min}_i, C^{max}_i]$ $=$ $[7.5, 12.5]$.

\subsection{Experimental Setup}
\label{subsec:setup}

To answer RQ1, RQ2, and RQ3 described in Section~\ref{subsec:rq}, we rely on case study data pertaining to ADCS, provided by LuxSpace, as well as the ICS and UAV subjects described in Section~\ref{subsec:casestudy}. To answer RQ4, we used 800 synthetic subjects  (see Section~\ref{subsec:synthetic}). We conducted four experiments, EXP1, EXP2, EXP3, and EXP4, as described below.

\noindent\textbf{EXP1.} To answer RQ1, we developed a baseline solution that estimates task WCETs based on random search (RS). The baseline replaces the GA in Phase 1 with RS and does not infer a safe border using logistic regression. Note that the baseline uses the same fitness function (see Section~\ref{subsec:phase1}) and also maintains the best population during search; however, it does not employ any genetic operators, i.e., crossover and mutation. The baseline solution also produces a labelled dataset $\vv{L}$ that contains tuples $(W,\ell)$ where $W$ is a set of task WCETs and $\ell$ is a label of $W$ indicating either safe or unsafe (see Section~\ref{subsec:phase1}). Given the labelled dataset, the baseline selects the best task WCETs that are safe and maximise the volume of the hyperbox they define. Specifically, the baseline finds a particular tuple $(W_s,\ell_s)$ in $\vv{L}$ that maximises the volume of the hyperbox defined by $W_s$ while satisfying the following condition: For all tuples $(W_x,\ell_x)$ in $\vv{L}$, the hyperbox defined by $W_x$ is contained in the hyperbox defined by $W_s$, $\ell_x$ $=$ \emph{safe}, and $\ell_s$ $=$ \emph{safe}. 

EXP1 compares the results obtained from executing SAFE and the baseline. For comparison, SAFE selects a best-size point, i.e., WCET ranges, on a safe border that maximises the volume of the hyperbox the point defines (see Section~\ref{subsec:phase2}). Given two solutions, i.e., estimated WCET ranges, obtained by SAFE and the baseline, EXP1 checks the schedulability of the two solutions using simulations. To do so, we ran simulations multiple times by varying task arrivals and task execution times within their estimated WCET ranges and checked whether there was a deadline miss in each simulation result.

\noindent\textbf{EXP2.} To answer RQ2, EXP2 compares our distance-based WCET sampling technique (described in Section~\ref{subsec:phase2}) with the naive random WCET sampling technique, for the second phase of SAFE. To this end, EXP2 first creates an initial training dataset by running the first phase of SAFE. EXP2 then relies on this initial training data for model refinement (Section~\ref{subsec:phase2}) by using both distance-based and naive random sampling. For comparison, EXP2 creates a test dataset by randomly sampling WCET values, which is independently created from the second phase of SAFE, and then compares the accuracy of the two sampling approaches in identifying safe WCET ranges for the test dataset.

\noindent\textbf{EXP3.} To answer RQ3, EXP3 computes precision values for SAFE, obtained from 10-fold cross-validation (see Section~\ref{subsec:phase2}), over each model refinement for the ADCS subject. We note that EXP3 focuses on ADCS to evaluate the practical usefulness of SAFE as we do not have any access to the engineers who have developed the other study subjects, i.e., ICS and UAV. In our study context, i.e., developing safety-critical systems, engineers require very high precision, i.e., ideally no false positives, (see Section~\ref{sec:approach}). Hence, EXP3 measures precision over model refinements to align with such practice. EXP3 then measures whether SAFE can compute safe WCET ranges within practical execution time and at an acceptable level of precision.

\noindent\textbf{EXP4.} To answer RQ4, EXP4 measures the execution time of SAFE with 800 synthetic systems. We use the task generation algorithm described in Section~\ref{subsec:synthetic} to create synthetic systems. In order to conduct controlled experiments to study correlations between the execution time of SAFE and a particular system parameter (e.g., number of tasks), we first create a baseline synthetic system by setting the parameters of Algorithm~\ref{alg:safewcet} as follows: (1)~We set the number of tasks $n$ to 20, the ratio of aperiodic tasks $\gamma$ to 0.45, and the maximum offset $\theta$ to 0. Note that these parameter values are the average values of the parameters in our industrial subjects. (2)~With regard to task periods, we set the range [$T^{min}$, $T^{max}$] of minimum and maximum periods to [10ms, 1s], which are common values in many real-time subjects~\cite{Baruah:11}. The granularity of task periods $g$ is set 10ms in order to increase realism as most of the task periods in our industrial subjects are multiples of 10ms. (3)~For the range factor to determine inter-arrival times for aperiodic tasks $\mu$, the number of WCET ranges $\omega$, the range factor to determine WCET ranges $\lambda$, and the target utilisation per processing core $u^t$, we assign $\mu$ = 0.25, $\omega$=2, $\lambda$ = 0.25, and $u^t$ = 0.9, respectively. We set these parameter values based on initial experiments to ensure that the executions of the synthetic systems examined in EXP4 sometimes violate their deadlines. Recall from Section~\ref{sec:approach} that SAFE relies on logistic regression and a labelled dataset, containing both safe (positive) and unsafe (negative) data instances. (4)~We set the number of processing cores $\epsilon$ to 1 as a baseline. (5)~For the simulation time of SafeScheduler (see Section~\ref{subsec:phase1}), we assign 30s in order to ensure that any aperiodic task arrives at least once and all possible arrivals of periodic tasks are analysed during that time.

Given the baseline system, we create several synthetic systems to be examined in EXP4 by varying the parameters' values as follows: (1)~number of tasks, $n$ $\in$ $\{5, 10, ..., 50\}$, (2)~ratio of aperiodic tasks, $\gamma$ $\in$ $\{0.05, 0.1, ..., 0.5\}$, (3)~range factor to determine inter-arrival times for aperiodic tasks, $\mu$ $\in$ $\{0.05, 0.1, ..., 0.5\}$, (4)~number of WCET ranges, $\omega$ $\in$ $\{1, 2, ..., 10\}$, (5)~range factor to determine WCET ranges, $\lambda$ $\in$ $\{0.05, 0.1, ..., 0.5\}$, (6)~maximum offset value $\theta$ $\in$ $\{200ms, 400ms,  ..., 2000ms\}$, (7)~number of processing cores, $\epsilon$ $\in$ $\{1, 2, ..., 10\}$, and (8)~simulation time, $\mathbf{t}$ $\in$ $\{30s, 1m, ..., 5m\}$. Note that we chose to study the effect of these parameters because they are controlled by engineers to design tasks in real-time systems. Simulation time $\mathbf{t}$ obviously impacts the execution time of SAFE as well. Resource dependencies are not controlled when generating synthetic systems as they do not impact SAFE's searching and learning (see Section~\ref{sec:approach}) but only simulations, which are investigated by varying simulation time. Due to the degree of randomness in our approach to generating synthetic systems (see Section~\ref{subsec:synthetic}), we create ten synthetic systems for each control parameter.

\subsection{Metrics}
\label{subsec:metrics}

We use the standard precision and recall metrics~\cite{Witten:11} to measure the accuracy in our experiments. To compute precision and recall in our context, for EXP2, we created a synthetic test dataset for each study subject containing tuples of WCET values and a flag indicating the presence or absence of deadline miss obtained from running SafeScheduler. Note that creating a test dataset by running an actual study subject with varying task WCETs is prohibitively expensive. We therefore used a set of task arrival sequences obtained from the first phase of SAFE for each subject as we aim at testing sequences of task arrivals which are more likely to violate their deadlines. We then ran SafeScheduler to simulate task executions for the set of task arrival sequences with randomly sampled WCET values. We note that WCET values were sampled within the restricted WCET ranges after the "handling imbalance" step in Algorithm~\ref{alg:safewcet}. Parts of the WCET ranges under which tasks are unlikely to be schedulable are therefore not considered when sampling.
For EXP3, we used 10-fold cross-validation based on the training dataset at each model refinement step (phase 2).

We define the precision and recall metrics as follows: (1)~precision $P = \mathit{TP} / (\mathit{TP} + \mathit{FP})$ and (2)~recall $R = \mathit{TP} / (\mathit{TP} + \mathit{FN})$, where $\mathit{TP}$, $\mathit{FP}$, and $\mathit{FN}$ denote the number of true positives, false positives, and false negatives, respectively. A true positive is a test instance (a set of WCET values) labelled as safe and correctly classified as such. A false positive is a test instance labelled as unsafe but incorrectly classified as safe. A false negative is a test instance labelled as safe but incorrectly classified as unsafe. We prioritise precision over recall as practitioners require (ideally) no false positives -- an unsafe instance with deadline misses is incorrectly classified as safe -- in the context of mission-critical, real-time satellite systems. For EXP2, precision and recall values are measured based on a synthetic test dataset. For EXP3, precision values are computed using collective sets of true positives and false positives obtained from 10-fold cross-validation at each model refinement.

Due to the inherent degree of randomness in SAFE, we repeat our experiments 50 times. For EXP1, we ran 40000 simulations to check the schedulability of the solutions obtained by SAFE and the baseline. To statistically compare our results, we use the non-parametric Mann-Whitney U-test~\cite{Mann:47} and Vargha and Delaney's $\hat{A}_{12}$ effect size~\citep{Vargha:00}. Mann-Whitney U-test determines whether two independent samples are likely or not to belong to the same distribution. We set the level of significance, $\alpha$, to 0.05. Vargha and Delaney’s $\hat{A}_{12}$ measures probabilistic superiority -- effect size -- between search algorithms. Two algorithms are considered to be equivalent when the value of $\hat{A}_{12}$ is 0.5.

\subsection{Implementation and Parameter Tuning}
\label{subsec:impl}

To implement the feature reduction step of Algorithm~\ref{alg:safewcet}, we used the random forest feature reduction~\cite{Breiman:01} as it has been successfully applied to high-dimensional data~\cite{Nguyen:15,Hideko:12}. For the stepwise regression step of Algorithm~\ref{alg:safewcet}, we used the stepwise AIC regression technique~\cite{Yamashita:07} which has been used in many applications~\cite{Zhang:16,May:16}. Recall from Section~\ref{subsec:phase2} that our distance-based sampling and best-size region recommendation require a numerical optimisation technique to find the nearest WCET sample and a maximum safe region size based on an inferred safe border. For such optimisations, we applied a standard numerical optimisation method, i.e., the Nelder-Mead method~\cite{Nelder:65}.

To compute the GA fitness, we set the number of SafeScheduler runs (Section~\ref{subsec:phase1}) for each solution  ($A$ in Section~\ref{subsec:phase1}) to 20. This number was chosen based on our initial experiments. We observed that 20 runs of SafeScheduler per solution $A$ keeps execution time under a reasonable threshold, i.e., $<$1.2m for all the subjects, and is sufficient to compute the fitness of SAFE. SafeScheduler schedules 34 tasks in ADCS for 1800s, 6 tasks in ICS for 150ms, and 16 tasks in UAV for 1500ms during which SafeScheduler advances its simulation clock by 0.1ms, 0.01ms, 0.01ms, respectively, for adequate precision.
We chose the time periods to ensure that all the tasks in each subject can be executed at least once.

For the GA search parameters, we set the population size to 10, the crossover rate to 0.7, and the mutation rate to 0.2, which are consistent with existing guidelines~\cite{Haupt:88}. We ran GA for 1000 iterations after which we observed that fitness reached a plateau in our initial experiments. Note that for the baseline comparison to be fair, we ran RS for 1500 iterations to ensure that the generated dataset $\vv{L}$ contained 30000 data instances, which are the same number of data instances obtained by SAFE.

Regarding the feature reduction step of Algorithm~\ref{alg:safewcet}, we set the random forest parameters as follows: (1)~the tree depth parameter is set to $\sqrt{|F|}$, where $|F|$ denotes the number of features, i.e., 26 WCET ranges in ADCS, 6 WCET ranges in ICS, and 16 WCET ranges in UAV, based on guidelines~\cite{Hastie:09}. (2)~The number of trees is set to 100 based on our initial experiments. We observed that learning more than 100 trees does not provide additional gains in terms of reducing the number of features.

Note that all the parameters mentioned above can probably be further tuned to improve the performance of SAFE. However, since with our current setting, we were able to convincingly and clearly support our conclusions, we do not report further experiments on tuning those parameters.

We ran our experiments over the high-performance computing cluster~\cite{Varrette:14} at the University of Luxembourg. To account for randomness, we repeated each run of SAFE 50 times for all the experiments. Each run of SAFE was executed on a different node of the cluster.  It took around 35h for us to create a synthetic test dataset with 50000 instances. When we set 1000 GA iterations for the first phase of SAFE and 10000 new WCET samples (100 refinements $\times$ 100 new WCET samples per refinement) for the second step of SAFE, each run of SAFE took at most 27.1h -- phase 1: 16.361h and phase 2: 10.74h. The running time is acceptable as SAFE can be executed offline in practice.

\subsection{Experiment Results}
\label{subsec:res}

\begin{table*}[t]

\caption{Comparing SAFE and our baseline method using (1)~the volumes of the hyperboxes that are defined by the best-size points computed by each method, (2)~the number of simulation runs that contain deadline misses out of total 40000 runs, and (3) the execution times of each method. The results are obtained from 50 runs of Safe and Baseline.}
\label{tbl:rq1 comparison}
\fontsize{8}{8}\selectfont
\begin{center}
\begin{tabularx}{\columnwidth}{l @{}Y@{}Y@{}Y@{}Y@{}Y@{}Y@{}Y@{}Y@{}Y@{}Y@{}Y@{}Y@{}Y@{}}
\toprule
 & \multicolumn{3}{c}{SAFE} & \multicolumn{3}{c}{Baseline} & \multicolumn{3}{c}{p-value} & \multicolumn{3}{c}{$\hat{A}_{12}$} \\
\cmidrule(lr){2-4} \cmidrule(lr){5-7} \cmidrule(lr){8-10} \cmidrule(lr){11-13}
 & ADCS & ICS & UAV & ADCS& ICS & UAV & ADCS & ICS & UAV & ADCS & ICS & UAV \\
\midrule 
\makecell[l]{Best-size\\ volumes} & \makecell{5.18e-11\\ ms$^{26}$} & \makecell{9.55e-01\\ ms$^{6}$} & \makecell{1.20e-02\\ ms$^{16}$} & \makecell{5.78e-12\\ ms$^{26}$}& \makecell{1.10e+00\\ ms$^{6}$}& \makecell{2.33e-04\\ ms$^{16}$} & 0.0000 & 0.0383 & 0.0000 & \textbf{1.0000} & 0.3796 & \textbf{1.0000} \\ 
\makecell[l]{Deadline\\ misses} & \textbf{0.00} & \textbf{5.42} & \textbf{1.36} & 114.92 & 13.2 & 32.32 & 0.0005 & 0.0023 & 0.0439 & \textbf{0.3900} & \textbf{0.3394} & \textbf{0.04084} \\ 
\makecell[l]{Execution\\ times} & \makecell{25.14\\ hours} & \makecell{1.37\\ hours} & \makecell{1.62\\ hours} & \makecell{24.29\\ hours} & \makecell{0.11\\ hours} & \makecell{0.31\\ hours} & 0.0000 & 0.0000 & 0.0000 & 0.8960 & 1.0000 & 1.0000 \\ 
\bottomrule
\end{tabularx}
\end{center}
\end{table*}

\noindent\textbf{RQ1.} Table~\ref{tbl:rq1 comparison} compares SAFE and our baseline method (Baseline) using the following three metrics: (1)~the volume of the hyperbox that is defined by the best-size point (see Sections~\ref{subsec:phase2} and \ref{subsec:setup}) computed by each method, i.e., SAFE and Baseline, (2)~the number of simulation runs, out of 40000 runs, that contain any deadline misses when tasks execute within their estimated WCET ranges defined by the best-size points, and (3)~the execution time of SAFE and Baseline to estimate WCET ranges. The results presented in the table are the mean values obtained from 50 runs of SAFE and Baseline for each of the three subjects. To enable accurate comparisons, we ran 40000 simulations of each execution of SAFE and Baseline, aiming at evaluating their estimated WCET ranges as described in Section~\ref{subsec:setup}. Statistical comparisons of the results obtained from 50 runs of SAFE and Baseline are summarized using p-values and $\hat{A}_{12}$ values as described in Section~\ref{subsec:metrics}.

As shown in Table~\ref{tbl:rq1 comparison}, compared to Baseline, SAFE provides more relaxed WCET ranges for ADCS and UAV. Note that the larger the hyperbox, the greater flexibility the engineers have in selecting appropriate WCET values. For ICS, however, Baseline finds a larger hyperbox than the best-size hyperbox produced by SAFE. This is likely due to the fact that ICS has a small number of tasks and is therefore much simpler than the other two subjects. Further, we recall that SAFE also provides engineers with a probability of deadline misses and trade-off relations between task WCETs based on an inferred logistic regression model (see Section~\ref{subsec:phase2}). We further discuss these benefits from a practitioner's perspective in RQ3.

EXP1 evaluates the estimated WCET ranges that are defined by the best-size points obtained from 50 runs of SAFE and Baseline for each subject. The estimated WCET ranges are examined through 40000 simulation runs by varying task arrivals and their execution times within the estimated WCET ranges. The ``Deadline misses'' row in Table~\ref{tbl:rq1 comparison} shows the mean number of simulation runs (out of 40000 runs) containing deadline misses. Across all the subjects, the differences between SAFE and Baseline are statistically significant as the p-values are less than 0.05. The $\hat{A}_{12}$ values show that SAFE is probabilistically superior to Baseline with respect to minimising the number of deadline misses.

Regarding the execution times of SAFE and Baseline, SAFE took more time than Baseline for all the subjects as shown in Table~\ref{tbl:rq1 comparison}. Estimating safe WCET ranges for ADCS requires the largest execution time (on average, 25.14h) compared to the other subjects. We note that such execution time is acceptable as SAFE can be executed offline in practice.

\begin{figure}[t]
\begin{center}
\includegraphics[width=0.4\linewidth]{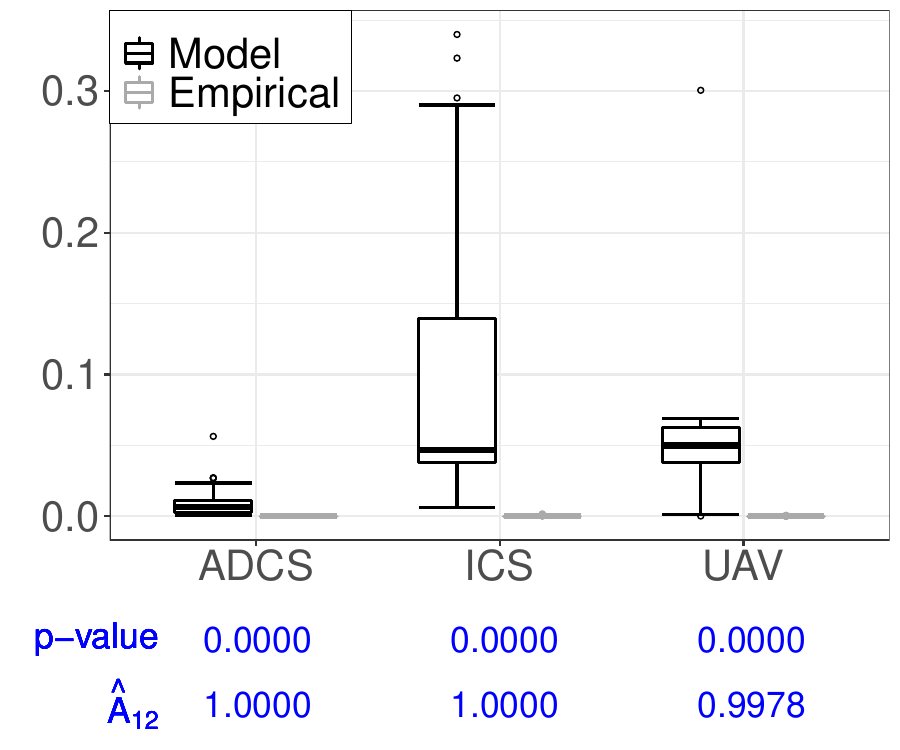}
\caption{Comparing probability values of deadline misses computed by SAFE and simulations for ADCS, ICS and UAV. SAFE uses a logistic regression model and selects a deadline miss probability to find the best-size WCET point. Empirical probability values of deadline misses are estimated based on 40000 simulations for each output, i.e., WCET ranges, of SAFE. The boxplots (20\%-50\%-75\%) show probability values obtained from 50 runs of SAFE (see Model) and 50 simulation-based evaluations (see Empirical), i.e., 50 $\times$ 40000 simulation runs.}
\label{fig:probability comparison}
\end{center}
\end{figure}

Figure~\ref{fig:probability comparison} shows probability distributions obtained from 50 runs of SAFE and simulations for ADCS, ICS, and UAV. As described in Section~\ref{subsec:phase2}, SAFE partitions the given WCET ranges into safe and unsafe sub-ranges using a safe WCET border that is defined by an inferred logistic regression model and a selected probability of deadline misses. In EXP1, SAFE selects a deadline miss probability that maximises the safe area under the safe border while ensuring that all the WCET points, i.e., sets of WCETs, classified as safe using the safe border are actually observed to be safe in the input dataset of logistic regression. The estimated WCET ranges, i.e., 50 best-size WCET points obtained by SAFE, are then evaluated through 40000 simulation runs for each WCET point by varying task arrivals and their execution time within their estimated WCET ranges. The empirical probability, i.e., relative frequency, of deadline misses is computed by the ratio of the number of simulation runs containing any deadline misses to the total number of simulation runs, i.e., 40000. The probability comparison depicted in Figure~\ref{fig:probability comparison} shows that the selected probability of deadline misses by SAFE is larger than the empirical probability computed by simulation-based evaluations. SAFE infers a logistic regression model, providing a probabilistic interpretation, based on a labelled dataset evaluated by worst-case task arrivals. The inferred logistic regression model, therefore, likely fits the system executions when task arrivals are worst with respect to maximising the magnitude of deadline misses. However, the empirical probability estimated through simulations is based on system executions when tasks randomly arrive within their inter-arrival time ranges. Hence, a logistic regression model obtained by SAFE enables more conservative probabilistic interpretations of the estimated WCET ranges than simulation-based evaluations for the WCET ranges. This implies that actual deadline miss probabilities tend to be lower than SAFE probability estimates, which is in practice a desirable property.

\begin{mdframed}[style=RQFrame]
\emph{The answer to {\bf RQ1} is that} SAFE significantly outperforms the baseline method with respect to minimising the number of deadline misses when using the estimated WCET ranges. Across our experiments, SAFE takes at most 27.1h (while the baseline takes 26.4h) to estimate the best-size WCET ranges. The execution time is acceptable as SAFE can be executed offline in practice.
\end{mdframed}

\begin{figure}[t]
\begin{center}
\subfloat[ADCS: Precision]{
    \includegraphics[width=0.4\columnwidth]{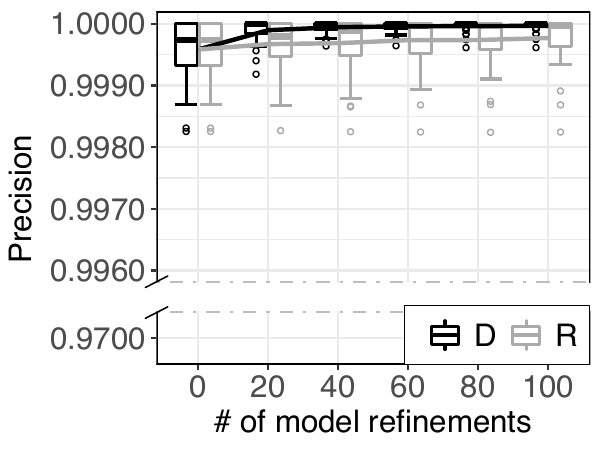}
    \label{fig:rq2 ADCS precision}
}%
\quad\quad
\subfloat[ADCS: Recall]{
    \includegraphics[width=0.4\columnwidth]{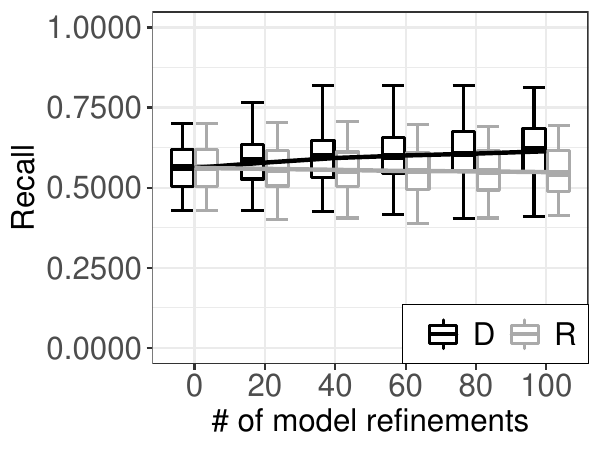}
    \label{fig:rq2 ADCS recall}
}%

\subfloat[ICS: Precision]{
    \includegraphics[width=0.4\columnwidth]{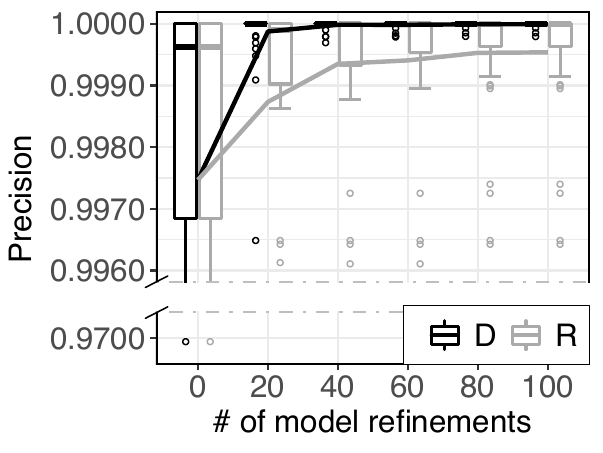}
    \label{fig:rq2 ICS precision}
}%
\quad\quad
\subfloat[ICS: Recall]{
    \includegraphics[width=0.4\columnwidth]{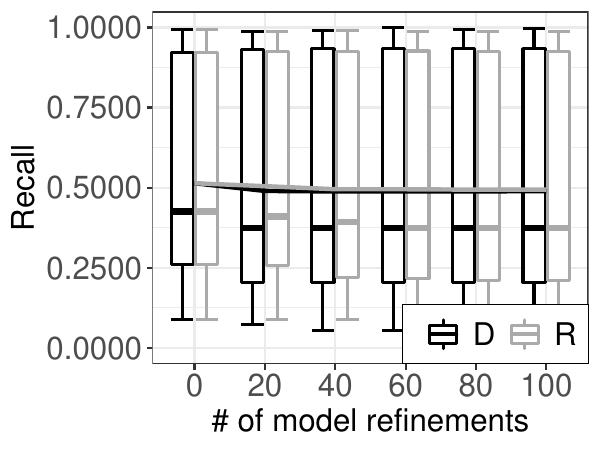}
    \label{fig:rq2 ICS recall}
}%

\subfloat[UAV: Precision]{
    \includegraphics[width=0.4\columnwidth]{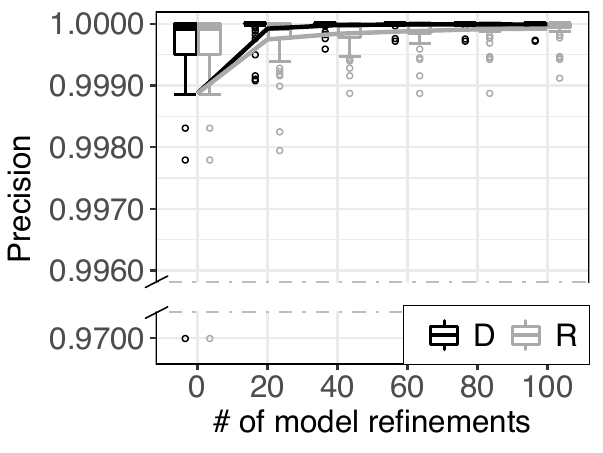}
    \label{fig:rq2 UAV precision}
}%
\quad\quad
\subfloat[UAV: Recall]{
    \includegraphics[width=0.4\columnwidth]{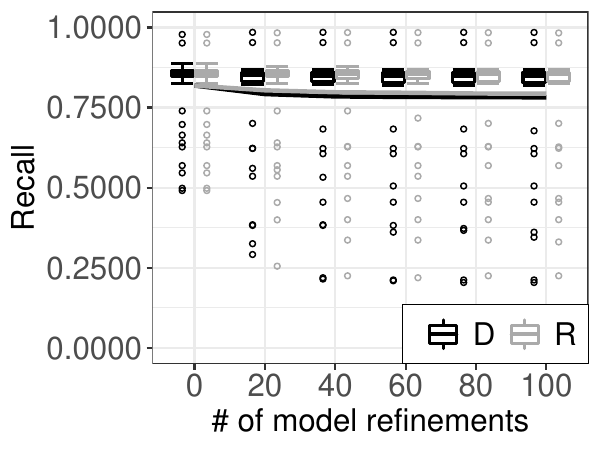}
    \label{fig:rq2 UAV recall}
}%
\caption{Distributions of precision and recall over 100 model refinements when SAFE employs either our distance-based sampling (D) or random sampling (R) for (a,b) ADCS, (c,d) ICS, and (e,f) UAV. The boxplots (25\%-50\%-75\%) show precision (a,c,e) and recall (b,d,f) values obtained from 50 runs of SAFE with each sampling strategy. The lines represent average trends.}
\label{fig:rq2}
\end{center}
\end{figure}

\noindent\textbf{RQ2.} Figure~\ref{fig:rq2} depicts distributions of precision (Figures~\ref{fig:rq2 ADCS precision}, \ref{fig:rq2 ICS precision}, and \ref{fig:rq2 UAV precision}) and recall (Figures~\ref{fig:rq2 ADCS recall}, \ref{fig:rq2 ICS recall}, and \ref{fig:rq2 UAV recall}) obtained from EXP2 with the ADCS, ICS, and UAV subjects. The boxplots in Figures~\ref{fig:rq2 ADCS precision}, \ref{fig:rq2 ICS precision}, and \ref{fig:rq2 UAV precision} (resp. Figures~\ref{fig:rq2 ADCS recall}, \ref{fig:rq2 ICS recall}, and \ref{fig:rq2 UAV recall}) show distributions (25\%-50\%-75\% quantiles) of precision (resp. recall) values obtained from 50 executions of SAFE with either  distance-based sampling (D) or simple random sampling (R). The solid lines represent the average trends of precision and recall value changes over 100 regression model refinements.

As shown in Figures~\ref{fig:rq2 ADCS precision}, \ref{fig:rq2 ICS precision}, and \ref{fig:rq2 UAV precision}, across over 100 model refinements, SAFE with D achieves higher precision values than those obtained by R for the three subjects. Also, Figures~\ref{fig:rq2 ADCS precision}, \ref{fig:rq2 ICS precision}, and \ref{fig:rq2 UAV precision} show that the variance of precision with D tends to be smaller than that of R. On average, D's precision converges toward 1 with model refinements; however, precision with R shows a markedly different trend that is not converging to 1, an important property in our context. Based on statistical comparisons, the difference in precision values between D and R becomes statistically significant after only 10, 4, and 11 model refinements, respectively, for ADCS, ICS, and UAV.

Regarding recall comparisons between D and R for ADCS, as shown in Figure~\ref{fig:rq2 ADCS recall}, D produces higher recall values over 100 model refinements than those of R. The difference in recall values between D and R becomes statistically significant after 36 model refinements. Regarding ICS (Figure~\ref{fig:rq2 ICS recall}) and UAV (Figure~\ref{fig:rq2 UAV recall}), their differences in  recall values for D and R are not statistically significant even after 100 model refinements. This may be explained by their much smaller number of tasks compared with ADCS. Across our experiments, for 100 model refinements, SAFE took, at most, 10.86h and 10.54h with D and R, respectively.

\begin{mdframed}[style=RQFrame]
	\emph{The answer to {\bf RQ2} is that}  SAFE with distance-based sampling significantly outperforms SAFE with random sampling in achieving higher precision. Only distance-based sampling can achieve a precision close to 1 within practical time, an important requirement in our context.
\end{mdframed}

\begin{figure}[t]
\begin{center}
\includegraphics[width=0.8\linewidth]{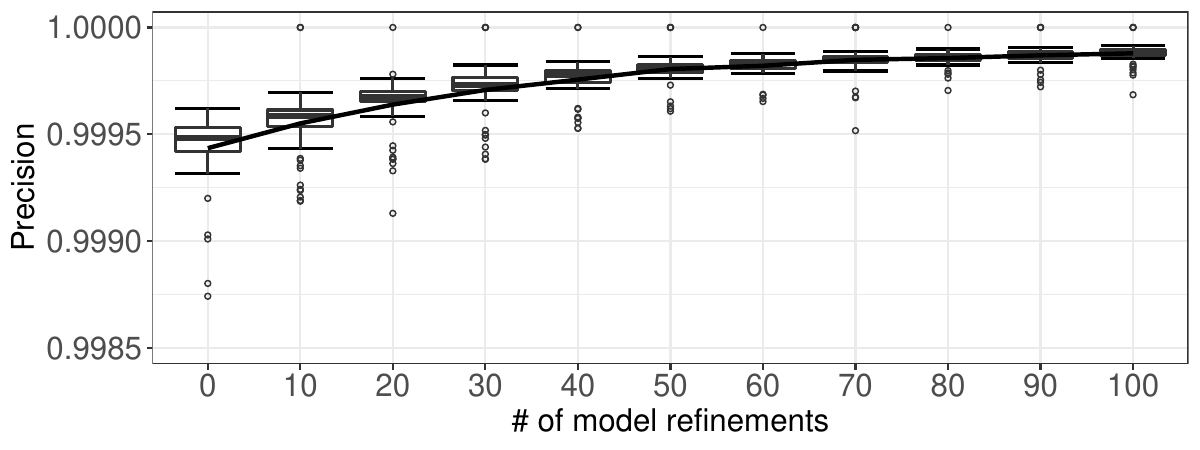}
\caption{Precision values computed from 10-fold cross validation at each model refinement for ADCS. The boxplots (25\%-50\%-75\%) show precision values obtained from 50 runs of SAFE. The line represents an average trend.}
\label{fig:kfold}
\end{center}
\end{figure}

\noindent\textbf{RQ3.} Figure~\ref{fig:kfold} shows precision values obtained from 10-fold cross-validation at each model refinement for the ADCS subject. Recall from Section~\ref{subsec:phase2} that SAFE stops model refinements once a precision value reaches a desired value. As shown in Figure~\ref{fig:kfold}, precision values tend to increase with additional WCET samples. Hence, practitioners are able to stop the model refinement procedure once precision reaches an acceptable level, e.g., $>$0.999. At 100 model refinement, SAFE reaches, on average, a precision of 0.99986. For EXP3, SAFE took, at most, 16.36h for phase 1 and 10.74h for phase 2.

As described in Section~\ref{subsec:phase2}, SAFE reduces the  dimensionality of the WCET space through a feature reduction technique based on random forest. The computed importance scores of each task's WCET in our dataset are as follows: 0.773 for T30, 0.093 for T33, 0.016 for T23, and $\leq$0.005 for the remaining 31 tasks. Based on a standard feature selection guideline~\cite{Hastie:09}, only the WCET values of two tasks, i.e., T30 and T33, are deemed to be important enough to retain as their score is higher than the average importance, i.e., 0.0385. Hence, SAFE computes safe WCET ranges of these two tasks in the next steps described in Algorithm~\ref{alg:safewcet}.

\begin{figure}[t]
\begin{center}
\includegraphics[width=0.8\linewidth]{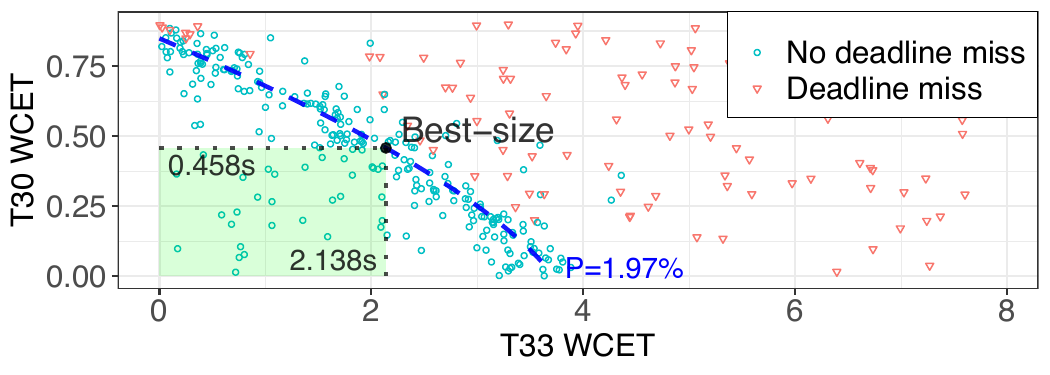}
\caption{An inferred safe border and best-size WCET regions for tasks T30 and T33. The safe border determines WCET ranges within which tasks are likely to be schedulable with a deadline miss probability of 1.97\%.}
\label{fig:bestsize}
\end{center}
\end{figure}

Figure~\ref{fig:bestsize} shows the inferred safe border which identifies safe WCET ranges within which all 34 tasks are schedulable with an estimated deadline miss probability of 1.97\%. 
Given the safe border, we found a best-size point which restricts the WCET ranges of T30 and T33 as follows: T30 [0.1ms, 458.0ms] and T33 [0.1ms, 2138.1ms]. We note that the initial estimated WCET ranges of the two tasks are as follows: T30 [0.1ms, 900.0ms] and T33 [0.1ms, 20000.0ms]. SAFE therefore resulted in safe WCET ranges representing a significant decrease of 49.11\% and 89.31\% of initial maximum WCET estimates, respectively. This information is therefore highly important and can be used to guide design and development.

\begin{mdframed}[style=RQFrame]
	\emph{The answer to {\bf RQ3} is that} SAFE helps compute safe WCET ranges that have a much lower maximum than practitioners' initial WCET estimates. Our case study showed that SAFE determined safe maximum WCET values that were only 51\% or less the original estimate. Further, these safe WCET ranges have a deadline miss probability of 1.97\% based on the inferred logistic regression model. More restricted ranges can be selected to reduce this probability. SAFE took, on average, 25.14h to compute such safe WCET regions, which is acceptable for offline analysis in practice.
\end{mdframed}

\begin{figure}[t]
\begin{center}
\subfloat[Number of tasks ($n$)]{
    \includegraphics[width=0.4\columnwidth]{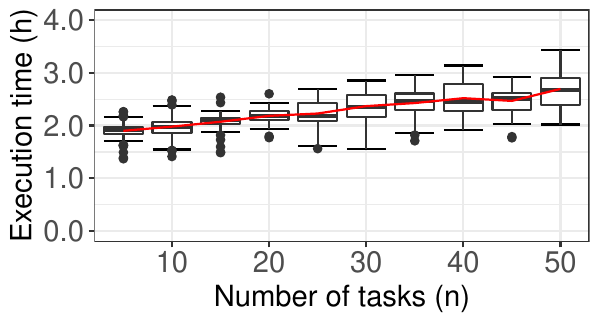}
    \label{fig:rq4 total w1}
}%
\quad\quad
\subfloat[Ratio of aperiodic tasks ($\gamma$)]{
    \includegraphics[width=0.4\columnwidth]{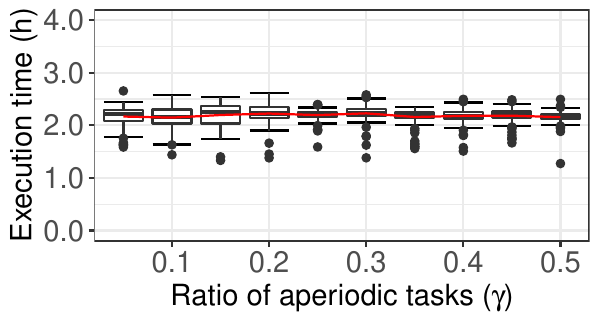}
    \label{fig:rq4 total w2}
}%

\subfloat[Range factor for inter-arrival times ($\mu$)]{
    \includegraphics[width=0.4\columnwidth]{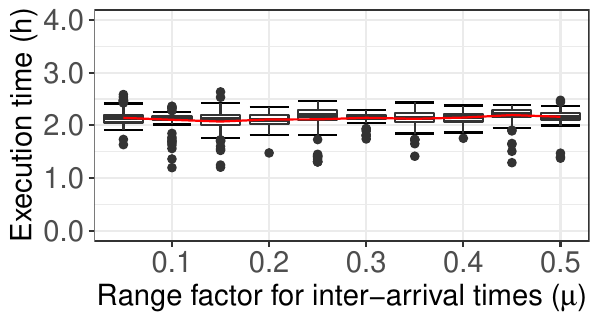}
    \label{fig:rq4 total w3}
}%
\quad\quad
\subfloat[Number of WCET ranges ($\omega$)]{
    \includegraphics[width=0.4\columnwidth]{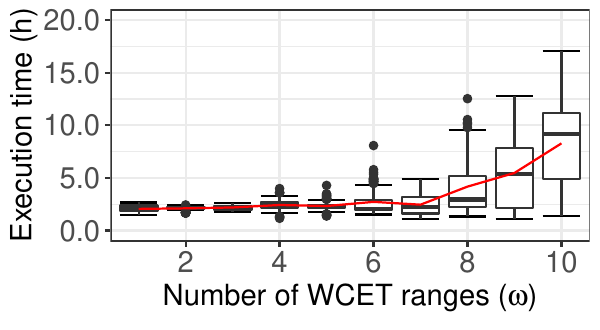}
    \label{fig:rq4 total w4}
}%

\subfloat[Range factor for WCET ranges ($\lambda$)]{
    \includegraphics[width=0.4\columnwidth]{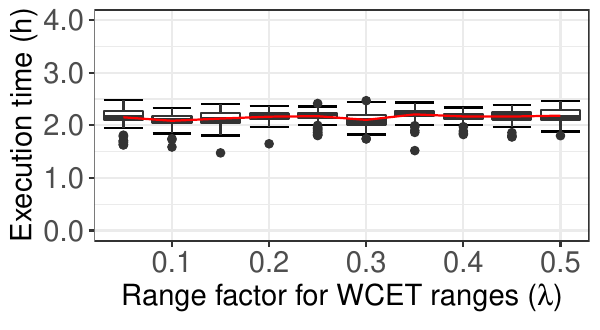}
    \label{fig:rq4 total w5}
}%
\quad\quad
\subfloat[Maximum offset value ($\theta$)]{
    \includegraphics[width=0.4\columnwidth]{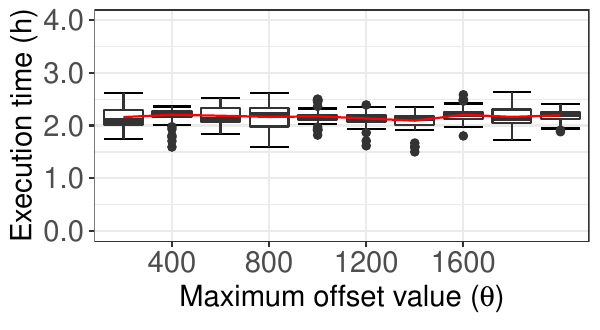}
    \label{fig:rq4 total w6}
}%

\subfloat[Number of processing cores ($\epsilon$)]{
    \includegraphics[width=0.4\columnwidth]{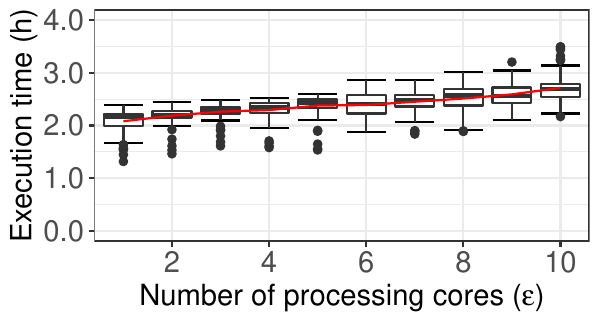}
    \label{fig:rq4 total w7}
}%
\quad\quad
\subfloat[Simulation time ($\mathbf{t}$)]{
    \includegraphics[width=0.4\columnwidth]{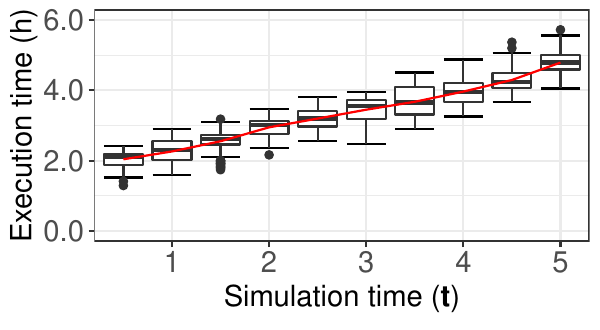}
    \label{fig:rq4 total w8}
}%
\caption{Execution times of SAFE when varying the values of the following parameters: (a)~number of tasks $n$, (b)~ratio of aperiodic tasks $\gamma$, (c)~range factor for inter-arrival times $\mu$, (d)~number of WCET ranges $\omega$, (e)~range factor for WCET ranges $\lambda$, (f)~maximum offset value $\theta$, (g)~number of processing cores $\epsilon$, and (h)~simulation time $\mathbf{t}$. The boxplots (20\%-50\%-75\%) show the distributions of execution time values obtained from 100 runs of SAFE, i.e., 10 runs for each of the 10 synthetic systems with the same configuration. The red line in each figure connects the mean values of the execution times of SAFE over the parameter values.}
\label{fig:rq4 total}
\end{center}
\end{figure}

\textbf{RQ4.} Figure~\ref{fig:rq4 total} shows the distributions (25\%-50\%-75\%) of execution times obtained from 10 $\times$ 10 runs of SAFE, i.e., 10 runs for each of the 10 synthetic systems with the same experimental setting (see Section~\ref{subsec:setup}). The red solid lines in Figure~\ref{fig:rq4 total} represent the mean value changes of the execution times of SAFE over varying values of the following control parameters: (a)~number of tasks $n$, (b)~ratio of aperiodic tasks $\gamma$, (c)~range factor for inter-arrival times $\mu$, (d)~number of WCET ranges $\omega$, (e)~range factor for WCET ranges $\lambda$, (f)~maximum offset value $\theta$, (g)~number of processing cores $\epsilon$, and (h)~simulation time $\mathbf{t}$. Note that all the experiments in EXP4 took at most 16.7h, which is acceptable as SAFE is an offline analysis technique.

As shown in Figures~\ref{fig:rq4 total w1}, \ref{fig:rq4 total w7}, and \ref{fig:rq4 total w8}, the execution time of SAFE is linear in the number of tasks $n$, the number of processing cores $\epsilon$, and the simulation time $\mathbf{t}$. However, regarding the ratio of aperiodic tasks $\gamma$ (Figure~\ref{fig:rq4 total w2}), the range factor for inter-arrival times $\mu$ (Figure~\ref{fig:rq4 total w3}), the range factor for WCET ranges $\lambda$ (Figure~\ref{fig:rq4 total w5}), the maximum offset value $\theta$ (Figure~\ref{fig:rq4 total w6}), the results indicate that they have no correlations with the execution time of SAFE. Therefore, we expect SAFE to scale well as the number of tasks, the number of processing cores, and the simulation time increase.

Figure~\ref{fig:rq4 total w4} shows that the execution time of SAFE is quadratically correlated with the number of WCET ranges $\omega$ in a system. Recall from Section~\ref{sec:approach} that the number of tasks characterising their WCETs as ranges (instead of point values) determine the size of a labelled dataset. Specifically, SAFE uses a second-order polynomial response surface model (RSM) to build a logistic regression model. RSM contains linear terms, quadratic terms, and 2-way interactions between linear terms (see Section~\ref{subsec:phase2}). Hence, the number of coefficients in an RSM to be inferred by logistic regression is the sum of the numbers of constants, linear terms, quadratic terms, and 2-way interactions in an RSM, i.e., $1 + \omega + \omega + \binom{\omega}{2}$ (see the RSM equation presented in Section~\ref{subsec:phase2}), which impacts the execution time of logistic regression dominating the execution time of SAFE. Note that the execution time of logistic regression is linear in the number of coefficients of a regression model (e.g., RSM)~\cite{David:86}. Hence, the execution time of logistic regression in SAFE is quadratically correlated with the number of WCET ranges as explained above. In addition, the results presented in Figure~\ref{fig:rq4 total w4} show that the magnitude of the execution time variation increases when the number of WCET ranges in a system increases. As written in Section~\ref{subsec:phase2}, SAFE employs a feature reduction technique and a stepwise regression technique to efficiently infer logistic regression models (see lines 1-4 of Algorithm~\ref{alg:safewcet}). The outputs of the two techniques depend on the number of tasks whose WCETs significantly impact deadline misses. Such outputs then impact the execution times of the following steps in Algorithm~\ref{alg:safewcet}: imbalance handling, sampling, and regression. Note that the synthetic systems generated by setting $\omega = 10$ are more diverse in terms of WCET ranges when compared to the systems created by setting $\omega = 1$.

In EXP4, SAFE analyses all tasks in a system. However, recall from Section~\ref{sec:approach} that SAFE provides the capability of selecting target tasks as engineers often need to focus on the most critical ones. In such cases, the execution time of SAFE can significantly decrease.

\begin{mdframed}[style=RQFrame]
	\emph{The answer to {\bf RQ4} is that} the execution time of SAFE is linear in the number of tasks, the number of processing cores, and the simulation time. However, the execution time of SAFE is quadratically correlated with the number of tasks whose WCETs are given as ranges. Across our experiments, SAFE took at most 17.1h, which is acceptable for offline analysis in practice.
\end{mdframed}

\noindent\textbf{Benefits from a practitioner's perspective.} Investigating practitioners' perceptions of the benefits of SAFE is necessary to adopt SAFE in practice. To do so, we draw on the qualitative reflections of three software engineers at LuxSpace, with whom we have been collaborating on this research. The reflections are based on the observations that the engineers made throughout their interactions with the researchers.

SAFE produces a set of worst-case sequences of task arrivals (see Section~\ref{subsec:phase1}). Engineers deemed them to be useful for further examinations by experts. The current practice is to use an analytical schedulability test~\cite{Liu:73} which proves whether or not a set of tasks are schedulable. Such an analytical technique typically does not provide additional information regarding possible deadline misses. In contrast, worst-case task arrivals and safe WCET ranges produced by SAFE offer insights to engineers regarding deadline miss scenarios and the conditions under which they happen.

Engineers noted that some tasks' WCET are inherently uncertain and that such uncertainty is hard to estimate based on expertise. Hence, their initial WCET estimates were very rough and conservative. Further, estimating what WCET sub-ranges are safe is even more difficult. Since SAFE estimates safe WCET ranges systematically with a probabilistic guarantee, the engineers deem SAFE to improve over existing practice. Also, SAFE allows engineers to choose system-specific safe WCET ranges from the (infinite) WCET ranges modeled by the safe border, rather than simply selecting the best-size WCET range automatically suggested by SAFE (Figure~\ref{fig:bestsize}). This flexibility allows engineers to perform domain specific trade-off analysis among possible WCET ranges and is useful in practice to support decision making with respect to their task design. 

Given the fact that we have not yet undertaken rigorous user studies, the benefits highlighted above are only suggestive but not conclusive. We believe the positive feedback obtained from LuxSpace and our industrial case study shows that SAFE is promising and worthy of further empirical research with human subjects.

\subsection{Threats to Validity}
\label{subsec:threats}

\textbf{Internal validity.} To ensure that our promising results cannot be attributed to the problem merely being simple, we compared SAFE with an alternative baseline using random search under identical parameter settings (see the RQ1 results in Section~\ref{subsec:res}). Phase 1 of SAFE can indeed be replaced with a random search, as we did, or even an exhaustive technique if the targeted system is small. However, there are no alternatives for Phase 2 -- our main contribution -- which infers safe WCET ranges and enables trade-off analysis. We present all the underlying parameters and provide our full evaluation package~\cite{Artifacts} to facilitate reproducibility. We mitigate potential biases and errors in our experiments by drawing on an industrial case study in collaboration with engineers at LuxSpace.

Recall from Section~\ref{subsec:res} that we compared probability values of deadline misses computed by SAFE and simulations. The results show that SAFE enables engineers to have more conservative probabilistic interpretations of the estimated WCET ranges than simulation-based evaluations for the WCET ranges. However, depending on the system's characteristics, e.g., hard real-time systems, engineers may need absolute guarantees for the WCET estimates. In such cases, once engineers find safe WCET ranges using SAFE, they can, in theory, obtain an absolute guarantee using exhaustive verification techniques, e.g., UPPAAL~\cite{Mikucionis:04}, on whether tasks always meet their deadlines for the given WCET ranges or not. Note that we performed an experiment using UPPAAL as it has often been used in the literature~\citep{Marius:10,Yu:10,Yalcinkaya:19}. We applied UPPAAL to verify whether ADCS tasks are schedulable for the given WCET values. However, our experiment results showed that UPPAAL was not able to complete the analysis task, even after five days of execution. Hence, engineers should therefore consider such scalability issues when applying exhaustive analysis techniques to complement SAFE. Since this UPPAAL evaluation is not the main focus of this article, we point the reader to the UPPAAL specification of ADCS available online~\citep{Artifacts}.

\textbf{External validity.} The main threat to external validity is that our results may not generalize to other contexts. We evaluated SAFE using early-stage WCET ranges estimated by practitioners at LuxSpace. However, SAFE can be applied at later development stages as well (1)~to test the schedulability of the underlying set of tasks of a system and (2)~to develop tasks under more precise constraints regarding safe WCETs. Future case studies covering the entire development process remain necessary for a more conclusive evaluation of SAFE. In addition, while motivated by ADCS (see Section~\ref{subsec:casestudy}) in the satellite domain, SAFE is designed to be generally applicable to other contexts.
To evaluate the usefulness of SAFE in other contexts, in addition to our motivating case study system, we applied SAFE to two industrial systems from different domains, having very different system characteristics such as resource dependencies and multiple processing cores. As we described in Section~\ref{subsec:casestudy}, however, none of the public study subjects provide initial estimates of their task WCETs as ranges, which are required by SAFE as input. Hence, we had to modify the study subjects to include WCET ranges in their task descriptions but attempted to minimise any potential biases and errors in the experiments by converting a point WCET value to a WCET range in a systematic and straightforward way. In Section~\ref{subsec:casestudy}, we described the converting method in detail. Also, we made the modified task descriptions available online~\cite{Artifacts}. However, the general usefulness of SAFE needs to be further assessed in other contexts and domains.

%% file: tex/relatedwork.tex
\section{Related Work}
\label{sec:relatedwork}

This section discusses and compares SAFE with related work in the areas of schedulability analysis, as well as testing and verification of real-time systems.

\noindent\textbf{Schedulability analysis} has been widely studied for real-time systems~\cite{Bernat:01,Bernat:02,Manolache:04,Axelsson:05,Xian:07,Bini:08,Cimatti:08,Gustafsson:09,Hansen:09,Muhuri:09,Maxim:13,Sun:13,Xu:15,Altenbernd:16,Hardy:16,Bonenfant:17,Bruggen:18,Andre:19,Pazzaglia:21,Shiva:12}. Among them, the most related research strands study uncertain execution times~\cite{Bini:08,Xian:07,Axelsson:05,Muhuri:09}, probability of deadline misses~\cite{Bruggen:18,Maxim:13,Manolache:04}, weakly hard deadlines~\cite{Bernat:01,Xu:15,Pazzaglia:21}, schedulability regions~\cite{Cimatti:08,Sun:13,Andre:19}, and WCET estimation~\cite{Bernat:02,Gustafsson:09,Hansen:09,Altenbernd:16,Hardy:16,Bonenfant:17} in the context of real-time task analysis.

Bini et al.~\cite{Bini:08} propose a theoretical sensitivity analysis method for real-time systems accounting for a set of periodic tasks and their uncertain execution times. Br{\"u}ggen et al.~\cite{Bruggen:18} present an analytical method to analyse a deadline miss probability of real-time tasks using probability density functions of approximated task execution times. In contrast to SAFE, most of these analytical approaches do not directly account for aperiodic tasks having variable arrival intervals; instead, they treat aperiodic tasks as periodic tasks using their minimum inter-arrival times as periods~\cite{Davis:11}. However, SAFE takes various task parameters, including irregular arrival times, into account without any unwarranted assumption. Also, our simulation-based approach enables engineers to explore different scheduling policies provided by real RTOS; however, these analytical methods are typically only valid for a specific conceptual scheduling policy model.

Bernat et al.~\cite{Bernat:01} introduce the concept of weakly hard real-time systems that can tolerate occasional deadline misses. They precisely define weakly hard deadline constraints, specifying a maximum number of deadlines that can be missed during a time window, and provide the theoretical analysis of the properties and relationships between tasks with the temporal constraints. Xu et al.~\cite{Xu:15} develop an algorithm that accounts for sporadic task overload when analysing the number of deadlines a task can miss in a given sequence of consecutive task arrivals. Pazzaglia et al.~\cite{Pazzaglia:21} present an extended weakly hard analysis method by accounting for additional uncertainties such as task offsets and release jitters. SAFE complements the above research strands on weakly hard real-time systems since, instead of analysing the number of deadlines a task can afford to miss over a time window, SAFE provides probabilistic guarantees for deadline misses based on logistic regression models inferred from search and simulation outputs.

Cimatti et al.~\cite{Cimatti:08} develop an approach that computes the regions of the task parameter values guaranteeing tasks are schedulable, i.e., schedulability regions. Their approach uses parametric timed automata and an SMT solver, i.e., NuSMT~\cite{Cavada:07}, and is applied to a system that contains two periodic tasks. Sun et al.~\cite{Sun:13} propose a method, named IMITATOR, that aims at computing schedulability regions. IMITATOR is based on model checking of parametric timed automata with stopwatches. They evaluate the method by applying it to two test-case systems that contain at most two free parameters, i.e., task execution times, that are defined as variables. Note that the other parameters, e.g., task periods and deadlines, are defined as fixed values. The results show that IMITATOR covers the entire parameter space but does not scale well with the size of the problem. Andr{\'{e}} et al.~\cite{Andre:19} developed a tool that translates a graphical specification of a real-time system to the input of IMITATOR such that it allows computation of some schedulability regions using IMITATOR. However, these methods that exhaustively search the problem space are often not amenable to analyse industrial systems in a scalable manner as they typically contain many tasks, different task types, complex task relationships, and multiple processing cores.

Hansen et al.~\cite{Hansen:09} present a measurement-based approach to estimate WCET and a probability of estimation failure. The measurement-based WCET estimation technique collects actual execution time samples and estimates WCETs using linear regression and a proposed analytical model. To our knowledge, most of the research strands regarding WCET estimation are developed for later development stages at which task implementations are available. Note that relatively few prior works aim at estimating WCET at an early design stage; however, these work strands still require access to source code, hardware, compilers, and program behaviour specifications~\cite{Gustafsson:09,Altenbernd:16,Bonenfant:17}. In contrast, SAFE uses as input estimated WCET ranges and then precisely restricts the WCET ranges within which tasks are schedulable with a selected deadline miss probability, by relying on a tailored genetic algorithm, simulation, feature reduction, a dedicated sampling strategy, and logistic regression.

\noindent\textbf{Testing and verification} are important to successfully develop safety-critical real-time systems~\cite{Mikucionis:04,Zander:08,Alesio:18,Alesio:12,Kwiatkowska:11,Briand:05}. Some prior studies employ model-based testing to generate and execute tests for real-time systems~\cite{Mikucionis:04,Zander:08,Alesio:18}. SAFE complements these prior studies by providing safe WCETs as objectives to engineers implementing and testing real-time tasks. Constraint programming and model checking have been applied to ensure that a system satisfies its time constraints~\cite{Alesio:12,Kwiatkowska:11}. These techniques may be useful to conclusively verify whether or not a WCET value is safe. However, such exhaustive techniques are not amenable to address the analysis problem addressed in this article, which requires the inference of safe WCET ranges. To our knowledge, SAFE is the first attempt to accurately estimate safe WCET ranges to prevent deadline misses with a given level of confidence and offer ways to achieve different trade-offs among tasks' WCET values.

%% file: tex/conclusion.tex
\section{Conclusion}
\label{sec:conclusion}

We developed SAFE, a two-phase approach applicable in early design stages, to precisely estimate safe WCET ranges within which real-time tasks are likely meet their deadlines with a high-level of confidence. SAFE uses a meta-heuristic search algorithm to generate worst-case sequences of task arrivals that maximise the magnitude of deadline misses, when they are possible. Based on the search results, SAFE uses a logistic regression model to infer safe WCET ranges within which tasks are highly likely to meet their deadlines, given a selected probability. SAFE is developed to be scalable by using a combination of techniques such as a genetic algorithm and simulation for the SAFE search (phase 1) and feature reduction, an effective sampling strategy, and polynomial logistic regression for the SAFE model refinement (phase 2). We evaluated SAFE on a mission-critical, real-time satellite system in collaboration with a satellite company as well as two industrial systems from different domains whose description was retrieved from the literature. The results indicate that SAFE is able to precisely compute safe WCET ranges for which deadline misses are highly unlikely, these ranges being much smaller than the WCET ranges initially estimated by engineers. 
Further, we evaluated the scalability of SAFE using a number of synthetic systems. The results indicate that SAFE scales to complex systems. Across the experiments on industrial and synthetic systems, SAFE took at most 27h, which is acceptable in practice as an offline analysis method. 

For future work, we plan to extend SAFE in the following directions: (1)~developing a real-time task modelling language to describe dependencies, constraints, behaviours of real-time tasks and to facilitate schedulability analysis and (2)~building a decision support system to recommend a schedulable solution if a set of tasks are not schedulable, e.g., priority re-assignments. In the long term, we would like to more conclusively validate the usefulness of SAFE by applying it to other case studies in different domains.

%% file: tex/acknowledgments.tex
\section*{Acknowledgment}

This project has received funding from the European Research Council (ERC) under the European Union's Horizon 2020 research and innovation programme (grant agreement No 694277), and NSERC of Canada under the Discovery and CRC programs. The experiments presented in this paper were carried out using the HPC facilities of the University of Luxembourg ~\cite{Varrette2014}{\small -- see \href{http://hpc.uni.lu}{hpc.uni.lu}}.